\documentclass{aa}
\usepackage{graphicx}
\usepackage{longtable}
\usepackage{lscape}
\usepackage{amsmath}
\usepackage{lmodern}

\begin{document}

\title{A new estimation of astrometric exoplanet detection limits in the habitable zone  around nearby stars.}

\titlerunning{new estimation of astrometric exoplanet detection limits}

\author{N. Meunier \inst{1}, A.-M. Lagrange   \inst{1,2}
  }
\authorrunning{Meunier et al.}

\institute{
Univ. Grenoble Alpes, CNRS, IPAG, F-38000 Grenoble, France \\
LESIA (UMR 8109), Observatoire de Paris, PSL Research University, CNRS, UMPC, Univ. Paris Diderot, 5 Place Jules Janssen, 92195 Meudon, France \\
     }

\offprints{N. Meunier}

\date{Received ; Accepted}

\abstract{Astrometry is less sensitive to stellar activity than the radial velocity technique when attempting to detect Earth mass planets in the habitable zone of solar-type stars. This is due to a smaller number of physical processes affecting the signal, and a larger ratio of the amplitude of the planetary signal to the stellar signal than with radial velocities. A few high-precision astrometric missions have therefore been proposed over the past two decades.}
{We aim to re-estimate the detection limits in astrometry for the nearby stars which are the main targets proposed for the THEIA astrometric mission, which is the most elaborate mission to search for planets, and to characterise its performance on the fitted parameters.  This analysis is performed for the 55 F-G-K stars in the THEIA sample. }
{We used realistic simulations of stellar activity and selected those that correspond  best to each star in terms of spectral type and average activity level. Then, we performed blind tests to estimate the performance.  }
{We find  worse detection limits compared to those previously obtained for that sample based on a careful analysis of the false positive rate, with values typically in the Earth-mass regime for most stars of the sample. The difference is attributed to the fact that we analysed full time series, adapted to each star in the sample, rather than using the expected solar jitter only. Although these detection limits have a relatively low signal-to-noise ratio, the fitted parameters have small uncertainties. }
{We confirm the low impact of stellar activity on exoplanet detectability for solar-type stars, although it plays a significant role for the closest stars such as $\alpha$ Cen A and B. We identify the best targets to be the stars with a close habitable zone.  However, for the few stars in the sample with a habitable zone corresponding to long periods, namely subgiants, the THEIA observational strategy is not well adapted and should prevent the detection of planets in the habitable zone, unless a longer mission can be proposed. }


\maketitle


\section{Introduction}

The THEIA mission \cite[][]{theia17} is a  high precision astrometric mission which was  proposed to  ESA, which aims at reaching several scientific objectives, including the detection of low mass planets around nearby stars. It may be proposed again in the future with new technological innovations \cite[][]{malbet21}. 
It follows several projects  which had not been selected \cite[see][for reviews]{malbet12,leger15,crouzier16,janson18}. We focus here on the exoplanet search in the habitable zone of those nearby stars. 
The impact of stellar activity on the astrometric signal has been taken into account in \cite{theia17} based on the jitter obtained for the Sun seen edge-on in one direction only
from \cite{lagrange11}. This study of the solar case followed several  approaches based on simple estimations of the stellar contribution \cite[][]{bastian05,reffert05,eriksson07,catanzarite08,lanza08}, and it was in good agreement with the results of the more complex solar reconstruction made by \cite{makarov09} with no plages and by \cite{makarov10} with a more complete model.

Our aim is to revisit the estimates made in \cite{theia17}, using recent developments  to estimate the contribution of stellar activity on the astrometric signal of the THEIA candidates more accurately. 
In \cite{meunier19}, hereafter Paper I, we extrapolated the solar simulations of \cite{borgniet15} to a larger range of spectral types (F6-K4) and activity levels, taking the complex distribution of spots and plages on the stellar surface into account. The impact of inclination was also studied.  
The complete set of astrometric time series at all inclinations was studied in \cite{meunier20}, hereafter Paper II, for a star at 10 pc. Inclination strongly impacts the signal in both directions, as also shown by \cite{sowmya21}. 
These time series allowed us, in Paper II, to compute detection limits as a function of spectral type for planets in the habitable zone of their host stars. Our objective here is to apply these tools to the stars selected as promising targets in \cite{theia17} by selecting the simulations corresponding the best to each target and to propose new detection limits based on these realistic activity simulations. We  selected the simulations corresponding to stars with the closest spectral type and with the closest levels of activity, and focus on the impact on the detectability of low mass exoplanets in the habitable zone.

The outline of the paper is the following. In Sect.~2, we present the targets and their properties as well as the simulation selection and properties. In Sect. 3, we compute the detection rates corresponding to the detection limits published in \cite{theia17} and our simulations to characterise them. Finally, in Sect.~4, we provide new detection limits based on our set of simulations, for planets in the habitable zones of the THEIA most promising targets. We conclude in Sect.~5.

\section{Methods}

We first describe the model used for the stellar contribution, due to spots and plages, and for the planet, and then the observational configuration. The target properties are introduced and we explain how we identified the simulations best suited to each target. Finally, we briefly present  their properties and in particular the signal-to-noise ratio (S/N), corresponding to the detection limits of \cite{theia17} and the simulations.

\subsection{Modelling stellar activity and planet}

We used the large amount of realistic simulations of astrometric and chromospheric time series,  described in detail in Paper I. They  are based on a complex solar-like distribution of spots and plages on the stellar surface, for FGK stars. The parameters cover a large range of stellar activity levels for different spectral types and relatively old main sequence stars. Two sets of time series have been produced according to the spot temperatures, one corresponding to a solar spot contrast ($\Delta T_{\rm spot1}$),  and the other to the upper limit from the sample of stars reported in \cite{berd05}, that is $\Delta T_{\rm spot2}$. Since this parameter directly impacts  the amplitude  of the variability, we consider both in the following. 
All time series were generated for 10  inclinations between 0$^\circ$ and 90$^\circ$, with a step of 10$^\circ$. 

These time series were analysed following a systematic approach in Paper II for stars at 10~pc. Detection rates for Earth-like planets and detection limits as a function of spectral type were obtained using different approaches.  In this paper, we adopt the observer point of view, and perform blind tests, in which the false positive level is determined from the false alarm probability (fap) using a classical bootstrap analysis on the time series including the planet. Appendixes C and D  also describe the results obtained with an approach 
 based on a frequential approach aiming at determining the power  corresponding to a false positive of 1~\% in the frequency domain we are interested in, before injecting any planet: the injection of a planet of a given mass at this period then allows to determine the corresponding detection rate for that mass and period.

The injected planets are assumed to be on circular orbits, and we considered systems with one planet only. 
 We do not expect the eccentricity to strongly impact the results if the temporal sampling is such that observations are spread over the whole duration, with no long gaps that may prevent one from characterising the eccentricity. If the system includes more massive planets, those will be detected and removed before searching for lower mass planets, and the residual is likely to include an additional noise contribution due to the uncertainty on the fit. The impact may be important in some cases, for example if the planet is not very massive compared to the Earth-like planets we are here focusing on, and has a long period compared to the temporal coverage of the observations: in this case, its orbit might not be well constrained. 
Their orbits are assumed to follow a distribution of inclinations with respect to the equatorial plane of the star similar to Paper II: we showed that the assumed distribution did not impact significantly the results.

\subsection{Observational strategy}

As in Paper II, we followed the strategy proposed in \cite{theia17}, i.e. 50 observations randomly distributed over the time-span of   3.5 years, and we used a noise level of 0.199 $\mu$as per observation for all single stars for the A component when observing a binary system.   We then  assumed that the exposure time will be adjusted according to the magnitude of the stars to provide this noise level. For B components of binary systems, the noise level is therefore naturally higher, and we scaled it according to the V magnitudes of the two stars.

\subsection{Targets and approach}

\subsubsection{Target choice}

\begin{figure*}
\includegraphics{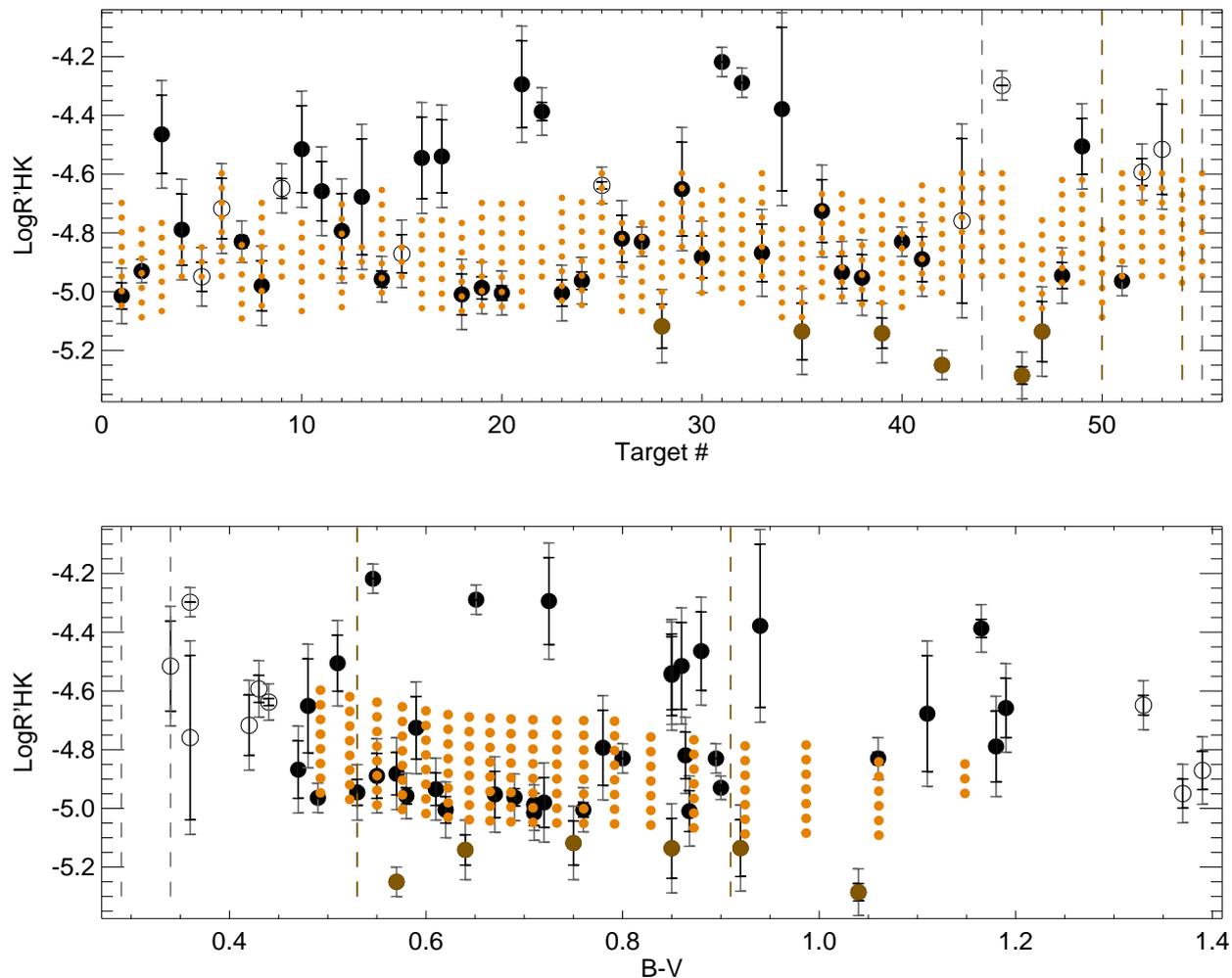}
\caption{
$\log R'_{HK}$ versus target number (upper panel) and versus B-V (lower panel) in black (main sequence) and brown (subgiants). Open symbols correspond to stars outside the simulation range in B-V.  Errorbars  indicate the range covered in the literature (black), and with the addition of a $\pm$0.05 uncertainty to the considered range (grey). The vertical dashed lines correspond to stars with no published $\log R'_{HK}$. The orange dots indicate the range covered by the simulations from Paper I (the actual range covers approximately $\pm$0.05 in $\log R'_{HK}$ given the different realisations and inclinations).   
}
\label{logrphk}
\end{figure*}

\begin{figure}
\includegraphics{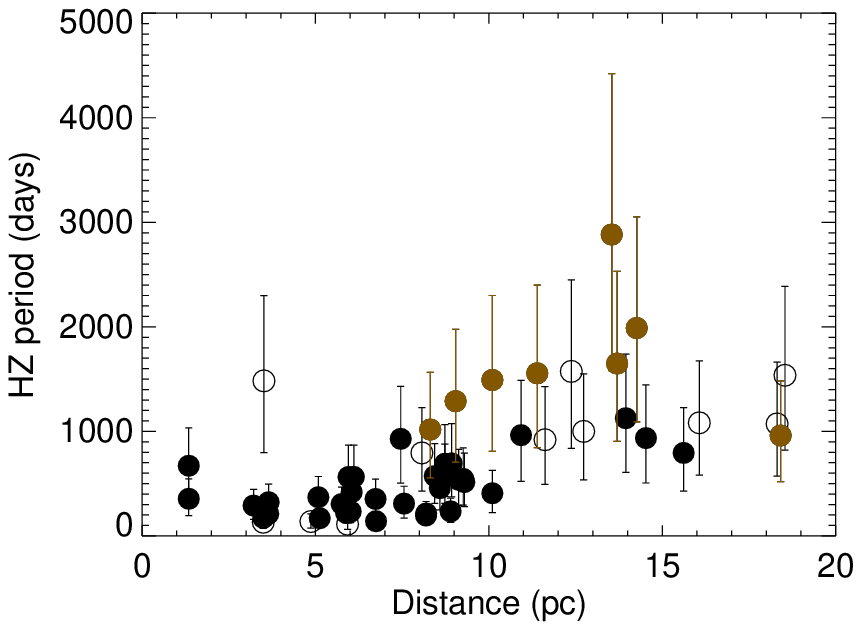}
\caption{Period in habitable zone for each star vs distance: the circles indicate the middle of the HZ, and the errobar symbol the extent of the HZ between PHZ$_{\rm in}$ and PHZ$_{\rm out}$. The colour and symbol codes are similar to those of Fig.~1: Subgiants are in brown, and open symbols correspond to stars outside the B-V range of our simulations (mostly F1-F5 stars). 
}
\label{phz}
\end{figure}

In this paper, we consider the stars listed in \cite{theia17}, which were considered to be the most promising targets for the  mission.  Most of them are F, G or K stars, which correspond to the spectral types considered in our simulations. A few of these stars (Table~\ref{tab_ts}) are A or M stars, and therefore fall well outside the range of the simulations made in Paper I: we therefore discarded these stars in the following.

\subsubsection{Target parameters}

\begin{table*}
\caption{Stellar-type statistics}
\label{tab_ts}
\begin{center}
\renewcommand{\footnoterule}{}  
\begin{tabular}{llll}
\hline
Stellar  & Number & Number of  & Status and comments \\
type &  of stars  &  B comp. &  \\
\hline
A & 2 & 0 &  Not treated here \\
F & 16 & 2 & 3 without $\log R'_{HK}$ information; 1 subgiant \\
G & 17 & 1 &  none  without $\log R'_{HK}$ information, 4 subgiants  \\
K & 22 & 9 &  1  without $\log R'_{HK}$ information, 1 subgiant \\
M & 6 & 1 &  Not treated here \\
\hline
\end{tabular}
\end{center}
\end{table*}

\begin{table*}
\caption{Adequation with simulation parameters}
\label{tab_compat}
\begin{center}
\renewcommand{\footnoterule}{}  
\begin{tabular}{llllll}
\hline
Spectral type & Total & Compatible  & Less     & More   & No published \\
                    &         & activity level & active   & active  &   $\log R'_{HK}$ \\
\hline
B-V in [0.49-1.15] & 44  & 29 & 3 & 10 & 2 (*) \\
B-V$<$0.49 & 8 & 5 & 0 & 1 & 2 \\
B-V$>$1.15 & 3 & 2 & 0 & 1 & 0 \\
\hline
Total & 55 & 36 & 3 (5) & 12 & 4 (2) \\
\hline
\end{tabular}
\end{center}
\tablefoot{(*) indicates two stars with no $\log R'_{HK}$ which we have considered as the most quiet stars (identified as subgiants). The total between parenthesis takes this into account. 
}
\end{table*}

The properties of our final sample of 55 FGK stars are shown in Table~\ref{tab_targets}. Their median distance is 8.3 pc, with values between 1.3 and 18.6 pc. 33~\% are single  (18 stars), while 45~\% are the A component of a binary system and 22~\% are the B component (subset of the A component of binary systems).

We  extracted from a large number of publications the activity level of those stars, defined by the usual $\log R'_{HK}$ index, which is a marker of the chromospheric emission. This index was also computed in the simulations presented in Paper I, and therefore can be used to establish a relationship between a given  star and the set of simulations in terms of average activity level. When only the S-index was available, we  converted it into a $\log R'_{HK}$ value based on the B-V from the CDS and the law from \cite{noyes84}.
In most cases, several values of $\log R'_{HK}$ are available for a given star. On the other hand, they are almost always published without an uncertainty on the measurement. Moreover, when published, they are small compared to the added uncertainty considered in this work (see Sect.~2.3.3).  We  retrieved them all, and kept the minimum and maximum values (Table~\ref{tab_targets}). This variability can be due to either intrinsic stellar variability, or uncertainties, and most likely to both. 
No $\log R'_{HK}$ was found for a few stars. 
Figure~\ref{logrphk} shows the range covered by the $\log R'_{HK}$  index for all stars for which at least one value was found, versus the target number (upper panel) and versus B-V (lower panel), allowing a comparison with the range covered by the simulations of Paper I (orange dots). The $\log R'_{HK}$ index ranges between -4.2 and -5.3.

We defined the habitable zone of each star in our sample using a procedure  similar to the one used in Paper II, with the exception that we directly used  the bolometric flux from \cite{theia17}  instead of using a T$_{\rm eff}$ law, since we also had to consider subgiants here. The definition of the habitable zone is based on the classical definition of \cite{kasting93}, who estimated where liquid water could be present on the surface based on luminosity effects only. In the following, we consider the middle of the habitable zone, PHZ$_{\rm mid}$, 
and the inner and outer sides, PHZ$_{\rm in}$ and PHZ$_{\rm out}$ respectively.
For subgiants (11\% of the FGK sample), the habitable zone is further away from the star than what was considered in Paper II (main sequence stars only), and therefore we expect a stronger planetary signal. On the other hand, the period is then in some cases longer than the temporal span of \cite{theia17}, so that we expect such a planet to be poorly characterised. Figure~\ref{phz} shows the habitable zone versus the distance of the stars in our sample. 
Most stars (70~\%) are below 10 pc, and mostly correspond to main sequence stars fully compatible with our simulations. On the other hand, subgiants and stars with a poor compatibility with our simulations are mostly between 10 and 20~pc in this sample (see below), which may bias the distance dependence.

\subsubsection{Correspondence with our simulation parameters}
\label{sect233}

We  kept all FGK stars from the sample of \cite{theia17}. However, a few of these stars are still outside the range of our simulation parameters, as shown in Fig.~\ref{logrphk}, mostly either in spectral type or because they are younger active stars. For example, the simulations covered a range in B-V between 0.49 and 1.15 (the spectral types were in the F6-K4 range), while the range covered by the target sample is between 0.29 and 1.39. Table~\ref{tab_compat} shows a summary of those configurations. 

For each star, we first selected the simulations corresponding to the closest B-V value. For F0-F5 stars, we used the F6 simulations. F5 stars probably behave in a similar way, while we may underestimate the signal of F0-F4 stars. Such stars will be flagged in the following (see Table~\ref{tab_var}). We applied the same procedure for K5-K7 stars, for which we used the simulations made for K4 stars, keeping in mind that we may slightly overestimate the signal. These flagged stars are also indicated with open circles in  Fig.~\ref{logrphk} and in subsequent plots.

In a second step, our objective is then, out of these simulations selected based on B-V, to extract those which are compatible with each target in terms of average activity level. For stars with published values of $\log R'_{HK}$, either a range in $\log R'_{HK}$ is available (44 stars, 80~\% of the sample) or a single level (7 stars, 13~\%). We then extended the range in $\log R'_{HK}$ (see Sect.~2.3.2) by $\pm$0.05  to take  the possible uncertainties \cite[][]{radick18} into account, as well as inclination effects (Paper I). We then selected the compatible simulations. Each astrometric time series (in two directions, the X-direction is  the direction of rotation, and the Y-direction is along the rotation axis) is then scaled to the distance of the target. 
A few stars are more active than those in our grid of simulations. For these stars, we used the most active simulations, again keeping in mind that the stellar activity level will be  underestimated.  On the other hand, subgiants are quiet given their $\log R'_{HK}$, and we considered the most quiet simulations in our set of series. 
When no value of $\log R'_{HK}$ was available (4 stars), we used the whole activity range of our simulations, except for the subgiants (selection of the lowest level of activity). 
This is summarised in Table~\ref{tab_compat}.  

Finally, we obtained between 81 and 648 compatible simulations (each computed for 10 inclinations) depending on the target, with a median value of 162 simulations. We also wish to produce different realisations due to the sampling, which differs from one realisation to the other. We therefore built for each target 1000 time series, corresponding to a random choice of simulations, inclinations, and temporal samplings. We considered both choices of $\Delta T_{\rm spot}$, which are most of the time analysed separately. A random gaussian noise is then added, according to Sect.~2.2.


\subsection{Global S/N corresponding to published detection limits}
\label{sect24}

\begin{figure*}
\includegraphics{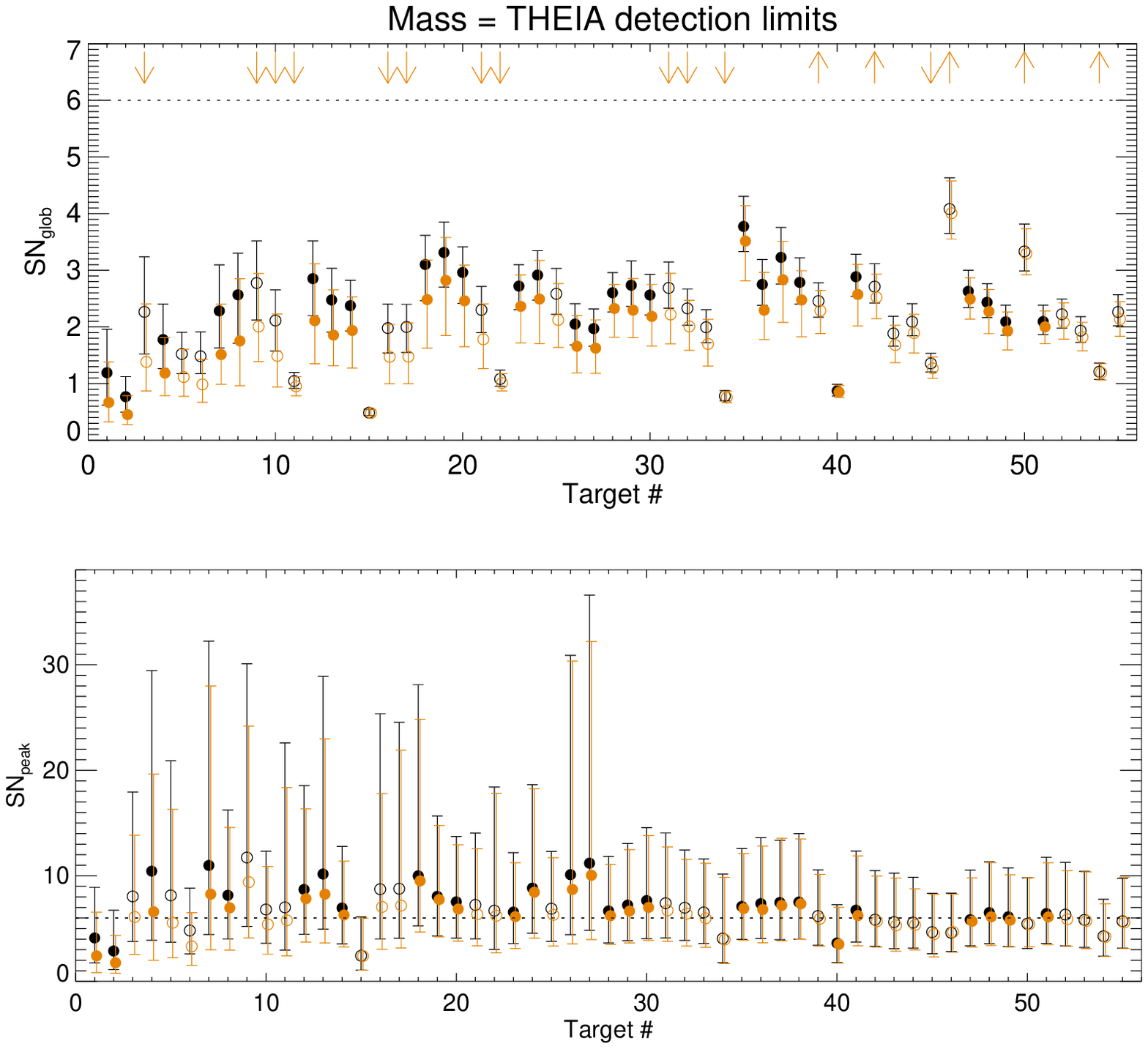}
\caption{SN$_{\rm glob}$ (upper panel) and SN$_{\rm peak}$ (lower panel, discussed in Appendix C.1)) vs target number, for planet masses at the detection limit of \cite{theia17}, for $\Delta T_{\rm spot1}$ (black) and $\Delta T_{\rm spot2}$ (orange). The horizontal dotted line corresponds to a S/N of 6. Circles correspond to the median value for each star (open circles correspond to stars outside the B-V range of our simulations), while the errorbar symbols correspond to the 5th and 95th percentiles. Upward arrows mean that the S/N are lower limits (quiet stars), while downward arrows correspond to upper limits (active stars). 
}
\label{sn}
\end{figure*}

\begin{figure*}
\includegraphics{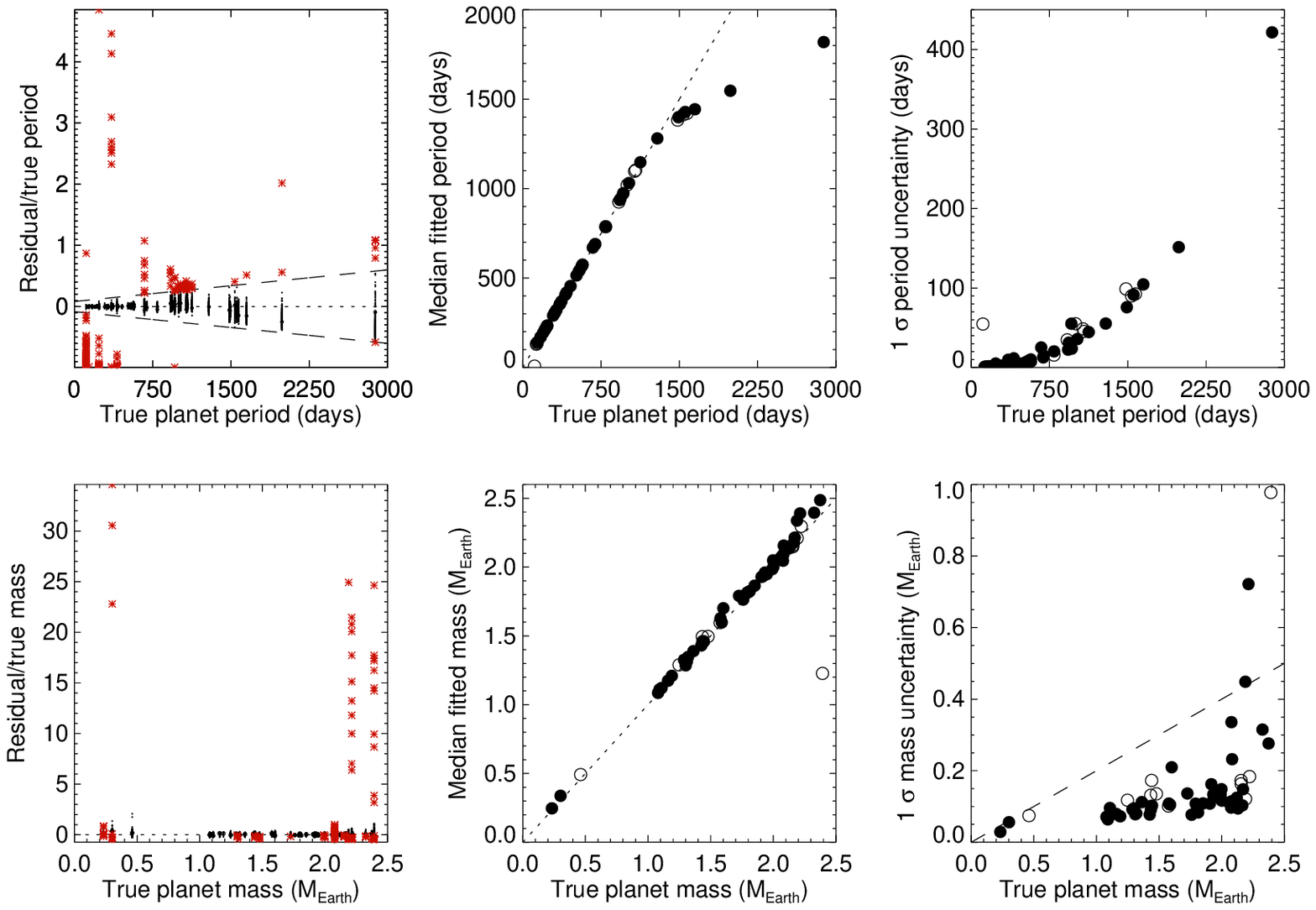}
\caption{Properties of fitted periods (upper panels) and fitted masses (lower panels) vs true values in the blind test for a planet mass at the detection limit of \cite{theia17}, for $\Delta T_{\rm spot1}$: relative residual for all peaks above the fap (left panels, false positives are in red, the dashed line on the period panel indicates the threshold between good detections and false positives we have considered), median values (middle panels, computed on all peaks above the fap, including the false positives, compared to the y=x dashed line), and 1-$\sigma$ uncertainties (right panels, the dashed line on the mass plot indicates a 20~\% uncertainty level). Open circles correspond to stars outside the B-V range of our simulations. 
}
\label{param}
\end{figure*}

\begin{figure*}
\includegraphics{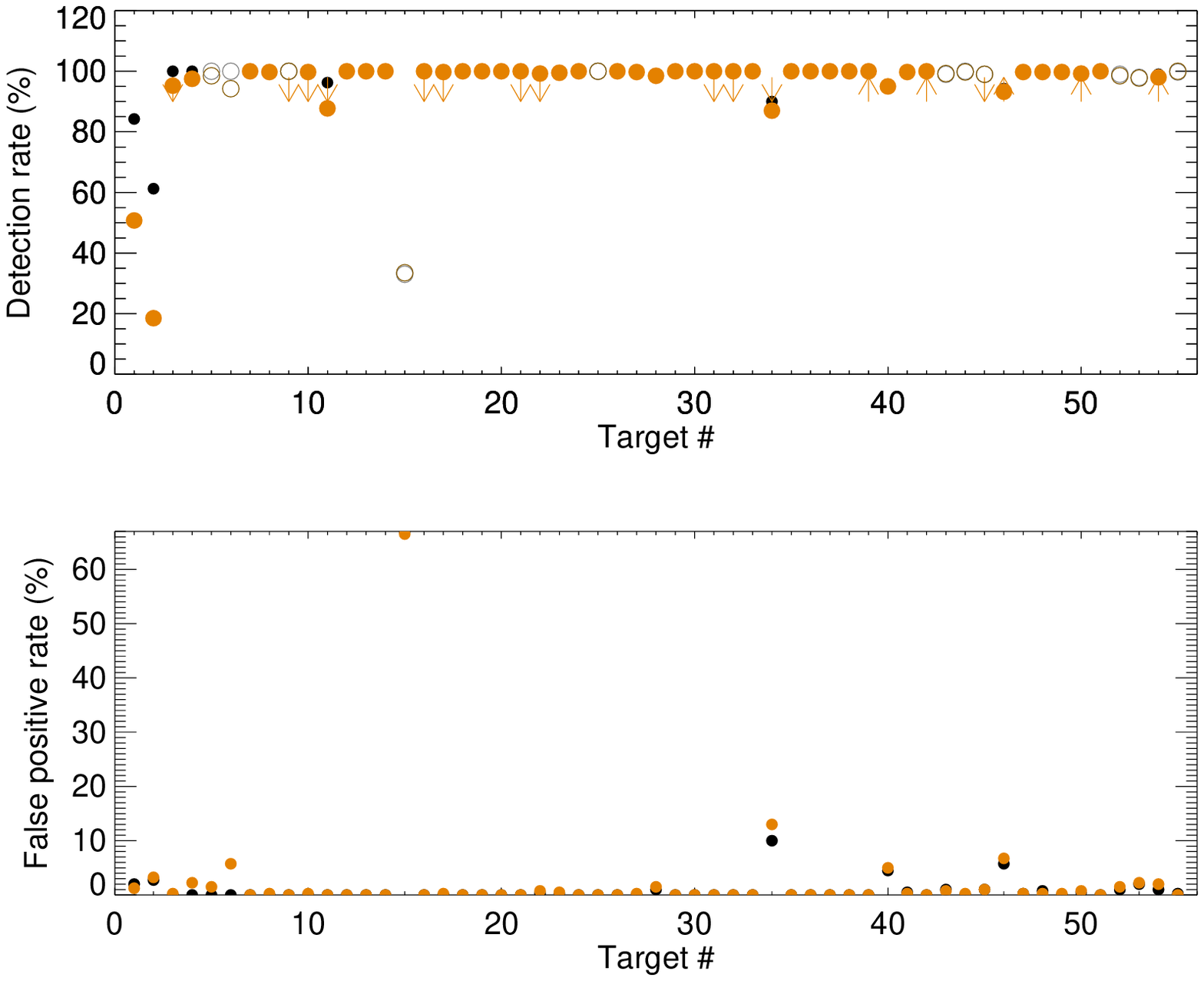}
\caption{Detection rates (upper panel) and false positive rates (lower panel) vs. target number for a planet mass at the 6-$\sigma$ (according to their analysis) detection limit of \cite{theia17}, for $\Delta T_{\rm spot1}$ (black) and $\Delta T_{\rm spot2}$ (orange), and for the blind test with a fap at 1~\%. Downward arrows mean that the detection rates are upper limits (active stars), while upward arrows correspond to lower limits (quiet stars, although most rates are already close to 100~\%). Open circles correspond to stars outside the B-V range of our simulations.
}
\label{tauxdet}
\end{figure*}

Before performing a detailed analysis of the time series, we  computed the typical signal-to-noise S/N corresponding to the detection limits published in \cite{theia17}. We recall that they were computed based on the solar value determined in \cite{lagrange11} in the X-direction, for a Sun seen edge-on and without scaling with the distance. We also considered the middle of the habitable zone we  estimated for each target.  The root mean square (herafter rms) due to the stellar contribution is studied in more detail in Appendix B for all targets. Here, we use the standard definition of the S/N, hereafter SN$_{\rm glob}$ (to distinguish it from another definition proposed in Appendix C.1), where the signal is considered to be $\alpha$, amplitude of the planetary signal. 
The noise was estimated from the rms of the time series including the stellar contribution and the noise of Sect.~2.2.

Figure~\ref{sn} (upper panel) shows the SN$_{\rm glob}$ for each target, where the indicated range of SN$_{\rm glob}$ for each target corresponds to the 5th and 95th percentiles.  The median SN$_{\rm glob}$ over the 55 stars is 2.3 and 1.9, for $\Delta T_{\rm spot1}$ and $\Delta T_{\rm spot2}$ respectively, i.e. well below the threshold of 6 indicated in \cite{theia17}. For $\Delta T_{\rm spot1}$ for example, median values vary between 0.5 and 4.1. We conclude that the published detection limits are associated with  a relatively small global S/N and not to a lower limit of 6. We provide another estimation of the performance in Sect.~3. For comparison, if we consider a similar planetary mass for all stars, a median SN$_{\rm glob}$ of 6 is reached for masses as high as 4.1~M$_{\rm Earth}$ and 5.2~M$_{\rm Earth}$, for $\Delta T_{\rm spot1}$ and $\Delta T_{\rm spot2}$ respectively. This is significantly above 1~M$_{\rm Earth}$, and above the detection limits of \cite{theia17} as well. 

The ratio between the rms due to activity and noise divided by the rms due to noise for all our targets is close to one for stars beyond $\sim$8~pc and $\sim$10 pc, for $\Delta T_{\rm spot1}$ and $\Delta T_{\rm spot2}$ respectively. This means that for stars at a  short distance  (typically 5 to 10 stars depending on whether we consider $\Delta T_{\rm spot1}$ and $\Delta T_{\rm spot2}$), stellar activity contributes significantly. A few stars have a similar contribution from the instrumental noise and stellar activity, and at larger distances, the noise is dominated by the instrumental noise (provided our assumption made in Sect.~2.2) and not by stellar activity.


\section{Detection rates corresponding to published detection limits}
\label{sect3}

In this section, we consider the detection limits provided by \cite{theia17} in the middle of the habitable zone and characterise them. We estimated the detection rate they correspond to for each star,  focusing on the blind test approach.
The results obtained with two other methods are described in Appendix C.
All values are provided in  Table~\ref{tab_rate}.

\begin{table}
\caption{Detection rates corresponding to published detection limits in \cite{theia17}}
\label{tab_rate2}
\begin{center}
\renewcommand{\footnoterule}{}  
\begin{tabular}{lllll}
\hline
Method  & Median & Minimum  & Maximum & Section\\
\hline
\multicolumn{5}{c}{$\Delta T_{\rm spot1}$} \\
\hline
Blind test & 100~\% & 33.0~\% & 100~\% & 3.1\\
SN$_{\rm peak}$ $>$ 6 &63.0~\%  & 5.1~\% & 91.4~\% & C.1\\
Theo. 1~\% & 100~\% & 13.0~\% & 100~\% & C.2\\
\hline
\multicolumn{5}{c}{$\Delta T_{\rm spot2}$} \\
\hline
Blind test & 100~\% & 18.5~\%& 100~\% & 3.1\\
SN$_{\rm peak}$ $>$ 6 & 53.1~\% & 1.2~\% & 85.7~\% & C.1\\
Theo. 1~\%  & 100~\% & 2.3~\% & 10~\%0 & C.2\\
\hline
\end{tabular}
\end{center}
\tablefoot{The blind test false positive level is estimated from a bootstrap analysis of the signal at the 1\% level, while the theoretical level (Theo.) is derived from the direct analysis of the simulation with no planet and no bootstrap. }
\end{table}

\subsection{Detection rates}

We performed blind tests on 400 time series per target (for a given $\Delta T_{\rm spot}$),  with the planet mass at the detection limit from \cite{theia17}, corresponding to 6-$\sigma$ (according to their analysis, but most likely to a lower level, see Sect.~2.4), and in the middle of the habitable zone. Following what was done in Paper II, for each time series, we computed the periodogram and the 1~\% fap level. If the highest peak is above the fap, we considered this peak to be a detection. If the peak is close to the planet period (the threshold is shown below), we considered it to be a good detection, otherwise it is a false positive. 

Figure~\ref{param} (left panels) shows the period (or mass) residual (difference between the fitted value and the true value) divided  by the true value, versus the true value. The dashed lines correspond to the thresholds we used to make the distinction between the good detections (in black) and the false positives (in red). We note that the false positives sometimes correspond to large masses, which is expected for the peaks at  low periods but similar signal amplitude. 

The middle panels of Figure~\ref{param} show the median of the  fitted periods and masses versus the true parameters for each target, for $\Delta T_{\rm spot1}$. There is a good correspondence, despite  a small bias on the mass and a departure from the true values for the highest periods, already observed in Paper II. The 1$\sigma$ uncertainties are shown in the left panels. The uncertainties on the periods are below 20 days for periods up to 1000 d (except for a noisy target, $\eta$ Cass. B). They naturally increase for larger periods. Most uncertainties on the masses are close to 10~\%, and below the 20~\% level, which is the targeted mass uncertainty for radial velocity follow-up of the PLAnetary Transits and Oscillations of stars mission (PLATO) detections: this level, which is not yet reachable in radial velocity for this type of stars \cite[e.g.][]{meunier22}, appears to be possible here for low planetary masses. For  $\Delta T_{\rm spot1}$ for example, only two stars (target 15, $\eta$ Cass. B, uncertainty 41~\% and target 34, $\gamma$ Lep B, uncertainty 33~\%) are above the 20~\% level (and target 40, zeta Herc B, is close to 20~\%).

Finally, the upper panel of Fig.~\ref{tauxdet} shows the detection rates (brown circles). They are  close to 100~\%, with the exception of $\alpha$ Cen A and B, which is discussed in Sect.~4, and $\eta$ Cass. B, which is noisy.  Table~\ref{tab_rate2}  summarises the median values over the 55 stars, which can be compared to the median values for the two methods described in Appendix C. If the false positive level is derived from the simulated time series with no planet (Appendix C.2), taking the frequency dependence of the signal into account, the derived detection rates are similar. On the other hand, the use of the S/N of the peak in the periodogram, SN$_{\rm peak}$ (Appendix C.1), leads to a median detection rate around 60\% only for a threshold of 6, and a large dispersion depending on the target. Table~\ref{tab_rate2}  provides the rates for each target.

\subsection{False positives}

We now consider the properties of the false positives. We identified the false positives from the blind tests following the definition above, i.e. peaks above the fap but with a period outside the true planetary period range (bottom panel of Fig.~\ref{tauxdet}). They are often at 0~\% or below the 1~\% expected level, except for a few stars around 5-10~\% (Procyon A, $\gamma$ Leporis B, $\zeta$ Herculis B, $\gamma$ Ceph B, i.e. either B components or stars with a long period of the habitable zone) and $\eta$ Cass. B because of its high level of noise, at 67~\%. The percentages are also slightly above 1~\% (in the 2-3~\% range) for $\alpha$ Cen A and B. For those stars, we also note that the average fap level is below the theoretical false positive level derived from the simulated time series alone in Appendix C.2 (while for all other stars it is at least twice higher, leading to a conservative estimation of the fap). This could explain a slightly larger amount of false positives compared to what is expected from the fap (1~\%). We conclude that for these two stars, due to their proximity, stellar activity effects may be above the considered level of noise.

\section{New detection limits in the whole habitable zone }
\label{sect4}

\begin{figure*}
\includegraphics{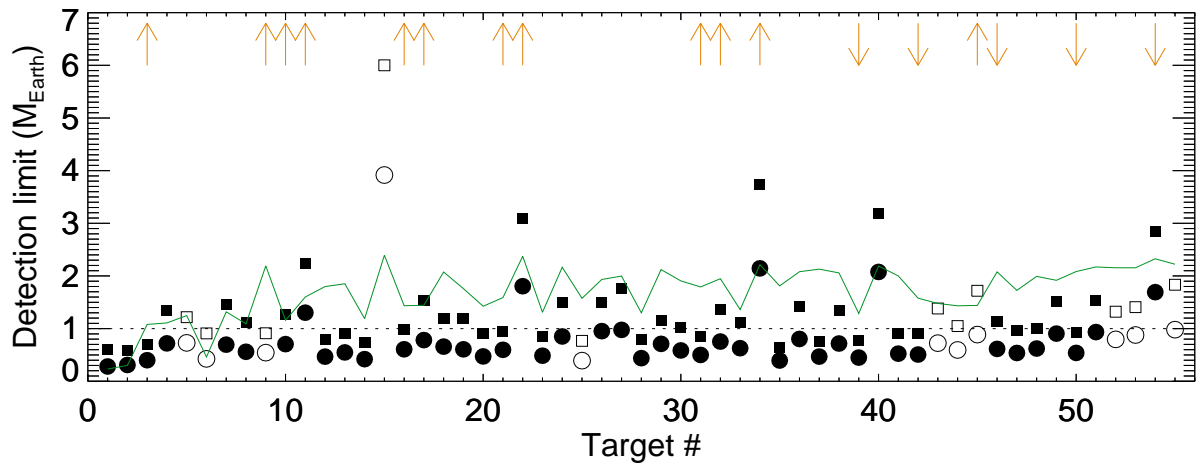}
\caption{
Detection limits vs. target number obtained with the blind test covering the whole habitable zone of each target and a random choice between $\Delta T_{\rm spot1}$ and $\Delta T_{\rm spot2}$. Circles are for a detection rate threshold of 50~\%, and squares for a detection rate threshold of 95~\%. Open circles correspond to stars outside the B-V range of our simulations. The  green line is the detection limit obtained in \cite{theia17}.
Downward arrows mean that the detection limits are upper limits (quiet stars), while upward arrows correspond lower limits (quiet stars). Detection limits of 10 M$_{\rm Earth}$ are lower limits (saturation of the detection rate curves, see Fig.~\ref{extaux}).
}
\label{limdet}
\end{figure*}

\begin{figure*}
\includegraphics{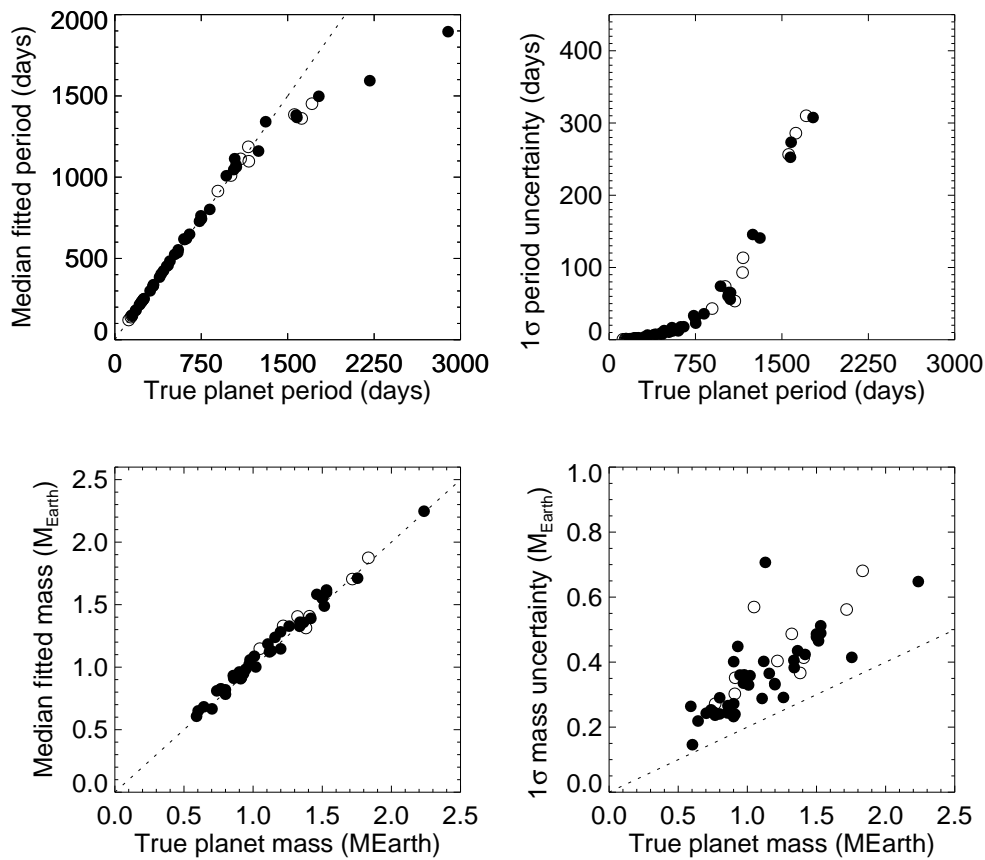}
\caption{
Properties of fitted periods (upper panels) and  masses (lower panels) vs true values for the masses the closest to the new detection limits from the blind test: median values (left panels, computed on all peaks above the fap, including the false positives, compared to the y=x dashed line), and 1-$\sigma$ uncertainties (right panels, the dashed line on the mass plot indicates a 20~\% uncertainty level). Open circles correspond to stars outside the B-V range of our simulations.
}
\label{par_new}
\end{figure*}


\begin{table}
\caption{New detection limits}
\label{tab_lim}
\begin{center}
\renewcommand{\footnoterule}{}  
\begin{tabular}{llll}
\hline
Method  & Median & Minimum  & Maximum \\
\hline
\multicolumn{4}{c}{Both $\Delta T_{\rm spot}$} \\
\hline
Blind test (50\%) & 0.28 & 0.62 & 3.92 \\
Blind test (95\%) & 0.59 & 1.13 & 6.00 \\
\hline
\multicolumn{4}{c}{$\Delta T_{\rm spot1}$} \\
\hline
SN$_{\rm peak}$ $>$ 6 (50\%) & 2.03 & 0.66 & (10)\\
Theo. 1\% (50\%) &  0.39 & 0.20 &  2.99 \\
Theo. 1\% (95\%) & 0.59 & 0.34 & 4.35 \\
\hline
\multicolumn{4}{c}{$\Delta T_{\rm spot2}$} \\
\hline
SN$_{\rm peak}$ $>$ 6 (50\%) & 2.48 & 1.20 & (10) \\
Theo. 1\% (50\%)  & 0.51 &  0.23 &  3.02 \\
Theo. 1\% (95\%)  & 0.75 & 0.38 & 4.49 \\
\hline
\end{tabular}
\end{center}
\tablefoot{Detection limits are in M$_{\rm Earth}$. The blind test false positive level is estimated from a bootstrap analysis of the signal at the 1\% level, while the theoretical level (Theo.) is derived from the direct analysis of the simulation with no planet and no bootstrap. The values between parenthesis are arbitrary (saturation of the detection rate below the 50~\% threshold for several stars). The 50~\% and 95~\% values in the method column correspond to the detection rate threshold used to compute the detection limits. }
\end{table}

We applied the same methods as in the previous section, but performed a loop on the mass in order to derive detection rates versus planet mass and then a detection limit corresponding to a well-defined detection rate. We considered planets in the whole habitable zone of their host stars. We focus on the detection limits obtained in the blind tests, and results with two additional approaches are shown in Appendix D. The new detection limits for all targets are given in Table~\ref{tab_limdet}

 We considered planets with a random period within the habitable zone and performed a loop on the mass between 0.05 and 10 M$_{\rm Earth}$.  A total of 200 realisations of the blind test were performed for each mass. Because these blind tests are time consuming, we choose randomly between the two spot contrasts at each realisation.
Fig.~\ref{limdet} shows the results for two thresholds on the detection rate, 50~\% and 95~\%. Table~\ref{tab_limdet} provides the resulting detection limits for all stars, and the median over the 55 stars is shown in Table~\ref{tab_lim} for comparison with the other methods. 
The detection limits based on the theoretical false positive levels derived from the direct analysis of the time series with no planets (Appendix D.2) are  comparable, but slightly lower than the blind test detection limits. On the other hand, the detection limits based on a threshold of 6 on SN$_{\rm peak}$ are significantly higher, so  the blind test detection limits correspond to relatively low SN$_{\rm peak}$ values.

Figure~\ref{par_new} shows the fitted versus true parameters in this blind test, for the mass closest to the detection limit at the 95\% detection rate level. We find a  good agreement, with the same departure at long periods already noted in the previous section. Masses being lower than the detection limits in \cite{theia17}, the uncertainties are slightly larger on the mass, typically around 30\%, which is still a good estimate. The uncertainty on the mass for such low-mass planets could probably be improved by considering more than 50 observations per star. 
Finally, the false positive rate is in most cases equal to 0 (out of the 200 realisations) or lower than 1\% (level chosen here for the fap level), except for a few stars with a long period (subgiants, F1 and F2 stars), with levels in the 2-4\% range.

\section{Discussion and conclusion}

We have performed simulations which take the impact of stellar activity on high precision time series into account with a good realism. We have applied our approach to all nearby F-G-K stars which have been identified as the most promising by \cite{theia17}. This allowed us to characterise the detections which could be made for planets at their detection limits, and to provide new detection limits for planets in the habitable zone of these stars. 

Our results differ from the detection limits in \cite{theia17} on several aspects. The blind test method  provides  lower detection limits, except for a handful of stars, showing that it is probably possible to reach lower masses.  
In particular, the median on the blind test for a 50\% detection rate is below 1~M$_{\rm Earth}$ and it is close to 1~M$_{\rm Earth}$ for a detection rate of 95\%, i.e. lower than the Super-Earth regime of the detection limits of \cite{theia17}. In such conditions, good uncertainties on the periods and masses can be achieved. This is much better than what can currently be achieved with the radial velocity technique \cite[]{dumusque17,meunier19b,meunier19e,meunier20b,meunier20c}.
The transit technique is currently biased towards planets close to their host stars and no planet in the habitable zone of solar-type stars have been detected so far: this is however a crucial objective of the PLATO mission, to be launched in 2026, which should allow one to detect such planets by the end of this decade, and will also provide key targets for future characteristion missions such as the Atmospheric Remote-sensing Infrared Exoplanet Large-survey (Ariel) or the Habitable Exoplanet Observatory (HabEX). The radial velocity technique, however, remains the only approach currently considered to estimate their mass, which is not possible for example using microlensing techniques (for which no follow-up is possible). 
Furthermore, the simple method based on SN$_{\rm peak}$ (with a threshold of 6) gives for many stars a good agreement with the published detection limits, but there are stars (mostly B components and subgiants) with a much higher detection limits based on this method. We note that  according to the standard definition of the S/N, which does not take into account the frequency behaviour of the signal, stellar activity should prevent the detection of planets in the Earth or Super-Earth regime. 

The new detection limits are therefore encouraging, since we can hope to detect low mass planets with a high precision astrometric mission without being affected by stellar activity, if the source of noise can be well taken into account. However, although they correspond to low false positive rates, they  therefore correspond to a low S/N in general: it will be important to understand well the different noise sources when analysing the data to validate these detections. 

We confirm that stellar activity does not play a significant role in general, except for the closest stars, and in particular $\alpha$ Cen A and B. For such stars, taking into account their activity when planning the observations (for example at time of cycle minimum) would improve the detection rates. Furthermore, the current strategy in \cite{theia17} is not well adapted for some of the B components (the level of noise is too high, but not due to stellar activity) and all subgiants in their target list (span of the observations too short to characterise planets in their habitable zone).  THEIA should not be able to secure detections in the habitable zone around those stars, unless the strategy is adapted with a significantly longer mission. The main sequence stars in the sample are therefore the most favourable targets.

\begin{acknowledgements}

We thank F. Malbet and A. Crouzier who provided the data from \cite{theia17}.
This work was supported by the "Programme National de Physique Stellaire" (PNPS) of CNRS/INSU co-funded by CEA and CNES.
This work was supported by the Programme National de Plan\'etologie (PNP) of CNRS/INSU, co-funded by CNES.

\end{acknowledgements}

\bibliographystyle{aa}
\bibliography{ms42702_bib}

\begin{thebibliography}{56}
\expandafter\ifx\csname natexlab\endcsname\relax\def\natexlab#1{#1}\fi

\bibitem[{{Aleo} {et~al.}(2017){Aleo}, {Sobotka}, \& {Ram{\'\i}rez}}]{aleo17}
{Aleo}, P.~D., {Sobotka}, A.~C., \& {Ram{\'\i}rez}, I. 2017, \apj, 846, 24

\bibitem[{{Allende Prieto} \& {Lambert}(1999)}]{allende99}
{Allende Prieto}, C. \& {Lambert}, D.~L. 1999, \aap, 352, 555

\bibitem[{{Baliunas} {et~al.}(1995){Baliunas}, {Donahue}, {Soon}, {Horne},
  {Frazer}, {Woodard-Eklund}, {Bradford}, {Rao}, {Wilson}, {Zhang}, {Bennett},
  {Briggs}, {Carroll}, {Duncan}, {Figueroa}, {Lanning}, {Misch}, {Mueller},
  {Noyes}, {Poppe}, {Porter}, {Robinson}, {Russell}, {Shelton}, {Soyumer},
  {Vaughan}, \& {Whitney}}]{baliunas95}
{Baliunas}, S.~L., {Donahue}, R.~A., {Soon}, W.~H., {et~al.} 1995, \apj, 438,
  269

\bibitem[{{Bastian} \& {Hefele}(2005)}]{bastian05}
{Bastian}, U. \& {Hefele}, H. 2005, in ESA Special Publication, Vol. 576, The
  Three-Dimensional Universe with Gaia, ed. C.~{Turon}, K.~S. {O'Flaherty}, \&
  M.~A.~C. {Perryman}, 215

\bibitem[{{Berdyugina}(2005)}]{berd05}
{Berdyugina}, S.~V. 2005, Living Reviews in Solar Physics, 2, 8

\bibitem[{{Borgniet} {et~al.}(2015){Borgniet}, {Meunier}, \&
  {Lagrange}}]{borgniet15}
{Borgniet}, S., {Meunier}, N., \& {Lagrange}, A.-M. 2015, \aap, 581, A133

\bibitem[{{Boro Saikia} {et~al.}(2018){Boro Saikia}, {Marvin}, {Jeffers},
  {Reiners}, {Cameron}, {Marsden}, {Petit}, {Warnecke}, \&
  {Yadav}}]{borosaikia18}
{Boro Saikia}, S., {Marvin}, C.~J., {Jeffers}, S.~V., {et~al.} 2018, \aap, 616,
  A108

\bibitem[{{Catanzarite} {et~al.}(2008){Catanzarite}, {Law}, \&
  {Shao}}]{catanzarite08}
{Catanzarite}, J., {Law}, N., \& {Shao}, M. 2008, in \procspie, Vol. 7013,
  Optical and Infrared Interferometry, 70132K

\bibitem[{{Cayrel de Strobel} {et~al.}(2001){Cayrel de Strobel}, {Soubiran}, \&
  {Ralite}}]{cayrel01}
{Cayrel de Strobel}, G., {Soubiran}, C., \& {Ralite}, N. 2001, \aap, 373, 159

\bibitem[{{Crouzier} {et~al.}(2016){Crouzier}, {Malbet}, {Henault},
  {L{\'e}ger}, {Cara}, {LeDuigou}, {Preis}, {Kern}, {Delboulbe}, {Martin},
  {Feautrier}, {Stadler}, {Lafrasse}, {Rochat}, {Ketchazo}, {Donati},
  {Doumayrou}, {Lagage}, {Shao}, {Goullioud}, {Nemati}, {Zhai}, {Behar},
  {Potin}, {Saint-Pe}, \& {Dupont}}]{crouzier16}
{Crouzier}, A., {Malbet}, F., {Henault}, F., {et~al.} 2016, \aap, 595, A108

\bibitem[{{Dumusque} {et~al.}(2017){Dumusque}, {Borsa}, {Damasso},
  {D{\'{\i}}az}, {Gregory}, {Hara}, {Hatzes}, {Rajpaul}, {Tuomi}, {Aigrain},
  {Anglada-Escud{\'e}}, {Bonomo}, {Bou{\'e}}, {Dauvergne}, {Frustagli},
  {Giacobbe}, {Haywood}, {Jones}, {Laskar}, {Pinamonti}, {Poretti}, {Rainer},
  {S{\'e}gransan}, {Sozzetti}, \& {Udry}}]{dumusque17}
{Dumusque}, X., {Borsa}, F., {Damasso}, M., {et~al.} 2017, \aap, 598, A133

\bibitem[{{Duncan} {et~al.}(1991){Duncan}, {Vaughan}, {Wilson}, {Preston},
  {Frazer}, {Lanning}, {Misch}, {Mueller}, {Soyumer}, {Woodard}, {Baliunas},
  {Noyes}, {Hartmann}, {Porter}, {Zwaan}, {Middelkoop}, {Rutten}, \&
  {Mihalas}}]{duncan91}
{Duncan}, D.~K., {Vaughan}, A.~H., {Wilson}, O.~C., {et~al.} 1991, \apjs, 76,
  383

\bibitem[{{Eriksson} \& {Lindegren}(2007)}]{eriksson07}
{Eriksson}, U. \& {Lindegren}, L. 2007, \aap, 476, 1389

\bibitem[{{Gaia Collaboration}(2018)}]{gaia18}
{Gaia Collaboration}. 2018, VizieR Online Data Catalog, I/345

\bibitem[{{Gray} {et~al.}(2006){Gray}, {Corbally}, {Garrison}, {McFadden},
  {Bubar}, {McGahee}, {O'Donoghue}, \& {Knox}}]{gray06}
{Gray}, R.~O., {Corbally}, C.~J., {Garrison}, R.~F., {et~al.} 2006, \aj, 132,
  161

\bibitem[{{Gray} {et~al.}(2003){Gray}, {Corbally}, {Garrison}, {McFadden}, \&
  {Robinson}}]{gray03}
{Gray}, R.~O., {Corbally}, C.~J., {Garrison}, R.~F., {McFadden}, M.~T., \&
  {Robinson}, P.~E. 2003, \aj, 126, 2048

\bibitem[{{Hall} {et~al.}(2007){Hall}, {Lockwood}, \& {Skiff}}]{hall07}
{Hall}, J.~C., {Lockwood}, G.~W., \& {Skiff}, B.~A. 2007, \aj, 133, 862

\bibitem[{{Henry} {et~al.}(1996){Henry}, {Soderblom}, {Donahue}, \&
  {Baliunas}}]{henry96}
{Henry}, T.~J., {Soderblom}, D.~R., {Donahue}, R.~A., \& {Baliunas}, S.~L.
  1996, \aj, 111

\bibitem[{{Holmberg} {et~al.}(2009){Holmberg}, {Nordstr{\"o}m}, \&
  {Andersen}}]{holmberg09}
{Holmberg}, J., {Nordstr{\"o}m}, B., \& {Andersen}, J. 2009, \aap, 501, 941

\bibitem[{{Houdebine} {et~al.}(2019){Houdebine}, {Mullan}, {Doyle}, {de La
  Vieuville}, {Butler}, \& {Paletou}}]{houdebine19}
{Houdebine}, {\'E}.~R., {Mullan}, D.~J., {Doyle}, J.~G., {et~al.} 2019, \aj,
  158, 56

\bibitem[{{Isaacson} \& {Fischer}(2010)}]{isaacson10}
{Isaacson}, H. \& {Fischer}, D. 2010, \apj, 725, 875

\bibitem[{{Janson} {et~al.}(2018){Janson}, {Brandeker}, {Boehm}, \&
  {Martins}}]{janson18}
{Janson}, M., {Brandeker}, A., {Boehm}, C., \& {Martins}, A.~K. 2018, {Future
  Astrometric Space Missions for Exoplanet Science}, 87

\bibitem[{{Jenkins} {et~al.}(2006){Jenkins}, {Jones}, {Tinney}, {Butler},
  {McCarthy}, {Marcy}, {Pinfield}, {Carter}, \& {Penny}}]{jenkins06}
{Jenkins}, J.~S., {Jones}, H.~R.~A., {Tinney}, C.~G., {et~al.} 2006, \mnras,
  372, 163

\bibitem[{{Kasting} {et~al.}(1993){Kasting}, {Whitmire}, \&
  {Reynolds}}]{kasting93}
{Kasting}, J.~F., {Whitmire}, D.~P., \& {Reynolds}, R.~T. 1993, \icarus, 101,
  108

\bibitem[{{Lagrange} {et~al.}(2011){Lagrange}, {Meunier}, {Desort}, \&
  {Malbet}}]{lagrange11}
{Lagrange}, A.-M., {Meunier}, N., {Desort}, M., \& {Malbet}, F. 2011, \aap,
  528, L9

\bibitem[{{Lanza} {et~al.}(2008){Lanza}, {De Martino}, \&
  {Rodon{\`o}}}]{lanza08}
{Lanza}, A.~F., {De Martino}, C., \& {Rodon{\`o}}, M. 2008, \na, 13, 77

\bibitem[{{L{\'e}ger} {et~al.}(2015){L{\'e}ger}, {Defr{\`e}re}, {Malbet},
  {Labadie}, \& {Absil}}]{leger15}
{L{\'e}ger}, A., {Defr{\`e}re}, D., {Malbet}, F., {Labadie}, L., \& {Absil}, O.
  2015, \apj, 808, 194

\bibitem[{{Lovis} {et~al.}(2011){Lovis}, {Dumusque}, {Santos}, {Bouchy},
  {Mayor}, {Pepe}, {Queloz}, {S{\'e}gransan}, \& {Udry}}]{lovis11b}
{Lovis}, C., {Dumusque}, X., {Santos}, N.~C., {et~al.} 2011, ArXiv e-prints
  1107.5325 [\eprint[arXiv]{1107.5325}]

\bibitem[{{Makarov} {et~al.}(2009){Makarov}, {Beichman}, {Catanzarite},
  {Fischer}, {Lebreton}, {Malbet}, \& {Shao}}]{makarov09}
{Makarov}, V.~V., {Beichman}, C.~A., {Catanzarite}, J.~H., {et~al.} 2009,
  \apjl, 707, L73

\bibitem[{{Makarov} {et~al.}(2010){Makarov}, {Parker}, \& {Ulrich}}]{makarov10}
{Makarov}, V.~V., {Parker}, D., \& {Ulrich}, R.~K. 2010, \apj, 717, 1202

\bibitem[{{Malbet} {et~al.}(2021){Malbet}, {Boehm}, {Krone-Martins}, {Amorim},
  {Anglada-Escud{\'e}}, {Brandeker}, {Courbin}, {En{\ss}lin}, {Falc{\~a}o},
  {Freese}, {Holl}, {Labadie}, {L{\'e}ger}, {Mamon}, {McArthur}, {Mora},
  {Shao}, {Sozzetti}, {Spolyar}, {Villaver}, {Abbas}, {Albertus}, {Alves},
  {Barnes}, {Bonomo}, {Bouy}, {Brown}, {Cardoso}, {Castellani}, {Chemin},
  {Clark}, {Correia}, {Crosta}, {Crouzier}, {Damasso}, {Darling}, {Davies},
  {Diaferio}, {Fortin}, {Fridlund}, {Gai}, {Garcia}, {Gnedin}, {Goobar},
  {Gordo}, {Goullioud}, {Hall}, {Hambly}, {Harrison}, {Hobbs}, {Holland},
  {H{\o}g}, {Jordi}, {Klioner}, {Lan{\c{c}}on}, {Laskar}, {Lattanzi}, {Le
  Poncin-Lafitte}, {Luri}, {Michalik}, {de Almeida}, {Mour{\~a}o}, {Moustakas},
  {Murray}, {Muterspaugh}, {Oertel}, {Ostorero}, {Portell}, {Prost},
  {Quirrenbach}, {Schneider}, {Scott}, {Siebert}, {Silva}, {Silva},
  {Th{\'e}bault}, {Tomsick}, {Traub}, {de Val-Borro}, {Valluri}, {Walton},
  {Watkins}, {White}, {Wyrzykowski}, {Wyse}, \& {Yamada}}]{malbet21}
{Malbet}, F., {Boehm}, C., {Krone-Martins}, A., {et~al.} 2021, Experimental
  Astronomy, 51, 845

\bibitem[{{Malbet} {et~al.}(2012){Malbet}, {L{\'e}ger}, {Shao}, {Goullioud},
  {Lagage}, {Brown}, {Cara}, {Durand}, {Eiroa}, {Feautrier}, {Jakobsson},
  {Hinglais}, {Kaltenegger}, {Labadie}, {Lagrange}, {Laskar}, {Liseau},
  {Lunine}, {Maldonado}, {Mercier}, {Mordasini}, {Queloz}, {Quirrenbach},
  {Sozzetti}, {Traub}, {Absil}, {Alibert}, {Andrei}, {Arenou}, {Beichman},
  {Chelli}, {Cockell}, {Duvert}, {Forveille}, {Garcia}, {Hobbs},
  {Krone-Martins}, {Lammer}, {Meunier}, {Minardi}, {Moitinho de Almeida},
  {Rambaux}, {Raymond}, {R{\"o}ttgering}, {Sahlmann}, {Schuller},
  {S{\'e}gransan}, {Selsis}, {Surdej}, {Villaver}, {White}, \&
  {Zinnecker}}]{malbet12}
{Malbet}, F., {L{\'e}ger}, A., {Shao}, M., {et~al.} 2012, Experimental
  Astronomy, 34, 385

\bibitem[{{Mamajek} \& {Hillenbrand}(2008)}]{mamajek08}
{Mamajek}, E.~E. \& {Hillenbrand}, L.~A. 2008, \apj, 687, 1264

\bibitem[{{Marfil} {et~al.}(2019){Marfil}, {Montes}, {Tabernero}, {Caballero},
  {Gonz{\'a}lez Hern{\'a}ndez}, {Kaminski}, {Sim{\'o}n-D{\'\i}az}, {Jeffers},
  {Quirrenbach}, {Amado}, {Ribas}, {Reiners}, {Seifert}, \& {CARMENES
  Consortium}}]{marfil19}
{Marfil}, E., {Montes}, D., {Tabernero}, H.~M., {et~al.} 2019, in Highlights on
  Spanish Astrophysics X, ed. B.~{Montesinos}, A.~{Asensio Ramos},
  F.~{Buitrago}, R.~{Sch{\"o}del}, E.~{Villaver}, S.~{P{\'e}rez-Hoyos}, \&
  I.~{Ord{\'o}{\~n}ez-Etxeberria}, 409--410

\bibitem[{{Meunier} \& {Lagrange}(2019{\natexlab{a}})}]{meunier19b}
{Meunier}, N. \& {Lagrange}, A.~M. 2019{\natexlab{a}}, \aap, 628, A125

\bibitem[{{Meunier} \& {Lagrange}(2019{\natexlab{b}})}]{meunier19e}
{Meunier}, N. \& {Lagrange}, A.~M. 2019{\natexlab{b}}, \aap, 625, L6

\bibitem[{{Meunier} \& {Lagrange}(2020{\natexlab{a}})}]{meunier20c}
{Meunier}, N. \& {Lagrange}, A.~M. 2020{\natexlab{a}}, \aap, 638, A54

\bibitem[{{Meunier} \& {Lagrange}(2020{\natexlab{b}})}]{meunier20b}
{Meunier}, N. \& {Lagrange}, A.~M. 2020{\natexlab{b}}, \aap, 642, A157

\bibitem[{{Meunier} \& {Lagrange}(2022)}]{meunier22}
{Meunier}, N. \& {Lagrange}, A.-M. 2022, in preparation

\bibitem[{{Meunier} {et~al.}(2020){Meunier}, {Lagrange}, \&
  {Borgniet}}]{meunier20}
{Meunier}, N., {Lagrange}, A.~M., \& {Borgniet}, S. 2020, \aap, 644, A77

\bibitem[{{Meunier} {et~al.}(2019){Meunier}, {Lagrange}, {Boulet}, \&
  {Borgniet}}]{meunier19}
{Meunier}, N., {Lagrange}, A.~M., {Boulet}, T., \& {Borgniet}, S. 2019, \aap,
  627, A56

\bibitem[{{Meunier} {et~al.}(2017){Meunier}, {Mignon}, \&
  {Lagrange}}]{meunier17b}
{Meunier}, N., {Mignon}, L., \& {Lagrange}, A.-M. 2017, \aap, 607, A124

\bibitem[{{Morel} {et~al.}(2001){Morel}, {Berthomieu}, {Provost}, \&
  {Th{\'e}venin}}]{morel01}
{Morel}, P., {Berthomieu}, G., {Provost}, J., \& {Th{\'e}venin}, F. 2001, \aap,
  379, 245

\bibitem[{{Noyes} {et~al.}(1984){Noyes}, {Hartmann}, {Baliunas}, {Duncan}, \&
  {Vaughan}}]{noyes84}
{Noyes}, R.~W., {Hartmann}, L.~W., {Baliunas}, S.~L., {Duncan}, D.~K., \&
  {Vaughan}, A.~H. 1984, \apj, 279, 763

\bibitem[{{Pace}(2013)}]{pace13}
{Pace}, G. 2013, \aap, 551, L8

\bibitem[{{Pourbaix} \& {Boffin}(2016)}]{pourbaix16}
{Pourbaix}, D. \& {Boffin}, H. M.~J. 2016, \aap, 586, A90

\bibitem[{{Radick} {et~al.}(2018){Radick}, {Lockwood}, {Henry}, {Hall}, \&
  {Pevtsov}}]{radick18}
{Radick}, R.~R., {Lockwood}, G.~W., {Henry}, G.~W., {Hall}, J.~C., \&
  {Pevtsov}, A.~A. 2018, \apj, 855, 75

\bibitem[{{Radick} {et~al.}(1998){Radick}, {Lockwood}, {Skiff}, \&
  {Baliunas}}]{radick98}
{Radick}, R.~R., {Lockwood}, G.~W., {Skiff}, B.~A., \& {Baliunas}, S.~L. 1998,
  \apjs, 118, 239

\bibitem[{{Reffert} {et~al.}(2005){Reffert}, {Launhardt}, {Hekker}, {Henning},
  {Queloz}, {Quirrenbach}, {S{\'e}gransan}, \& {Setiawan}}]{reffert05}
{Reffert}, S., {Launhardt}, R., {Hekker}, S., {et~al.} 2005, in Astronomical
  Society of the Pacific Conference Series, Vol. 338, Astrometry in the Age of
  the Next Generation of Large Telescopes, ed. P.~K. {Seidelmann} \& A.~K.~B.
  {Monet}, 81

\bibitem[{{Schr{\"o}der} {et~al.}(2009){Schr{\"o}der}, {Reiners}, \&
  {Schmitt}}]{schroder09}
{Schr{\"o}der}, C., {Reiners}, A., \& {Schmitt}, J.~H.~M.~M. 2009, \aap, 493,
  1099

\bibitem[{{Sowmya} {et~al.}(2021){Sowmya}, {N{\`e}mec}, {Shapiro},
  {I{\c{s}}{\i}k}, {Witzke}, {Mints}, {Krivova}, \& {Solanki}}]{sowmya21}
{Sowmya}, K., {N{\`e}mec}, N.~E., {Shapiro}, A.~I., {et~al.} 2021, \apj, 919,
  94

\bibitem[{{The Theia Collaboration} {et~al.}(2017){The Theia Collaboration},
  {Boehm}, {Krone-Martins}, {Amorim}, {Anglada-Escude}, {Brandeker}, {Courbin},
  {Ensslin}, {Falcao}, {Freese}, {Holl}, {Labadie}, {Leger}, {Malbet}, {Mamon},
  {McArthur}, {Mora}, {Shao}, {Sozzetti}, {Spolyar}, {Villaver}, {Albertus},
  {Bertone}, {Bouy}, {Boylan-Kolchin}, {Brown}, {Brown}, {Cardoso}, {Chemin},
  {Claudi}, {Correia}, {Crosta}, {Crouzier}, {Cyr-Racine}, {Damasso}, {da
  Silva}, {Davies}, {Das}, {Dayal}, {de Val-Borro}, {Diaferio}, {Erickcek},
  {Fairbairn}, {Fortin}, {Fridlund}, {Garcia}, {Gnedin}, {Goobar}, {Gordo},
  {Goullioud}, {Hambly}, {Hara}, {Hobbs}, {Hog}, {Holland}, {Ibata}, {Jordi},
  {Klioner}, {Kopeikin}, {Lacroix}, {Laskar}, {Le Poncin-Lafitte}, {Luri},
  {Majumdar}, {Makarov}, {Massey}, {Mennesson}, {Michalik}, {Moitinho de
  Almeida}, {Mourao}, {Moustakas}, {Murray}, {Muterspaugh}, {Oertel},
  {Ostorero}, {Perez-Garcia}, {Platais}, {de Mora}, {Quirrenbach}, {Randall},
  {Read}, {Regos}, {Rory}, {Rybicki}, {Scott}, {Schneider}, {Scholtz},
  {Siebert}, {Tereno}, {Tomsick}, {Traub}, {Valluri}, {Walker}, {Walton},
  {Watkins}, {White}, {Evans}, {Wyrzykowski}, \& {Wyse}}]{theia17}
{The Theia Collaboration}, {Boehm}, C., {Krone-Martins}, A., {et~al.} 2017,
  arXiv:1707.01348 [\eprint[arXiv]{1707.01348}]

\bibitem[{{Valenti} \& {Fischer}(2005)}]{valenti05}
{Valenti}, J.~A. \& {Fischer}, D.~A. 2005, \apjs, 159, 141

\bibitem[{{van Leeuwen}(2007)}]{vanleeuwen07}
{van Leeuwen}, F. 2007, \aap, 474, 653

\bibitem[{{White} {et~al.}(2020){White}, {Tapia-V{\'a}zquez}, {Hughes},
  {Mo{\'o}r}, {Matthews}, {Wilner}, {Aufdenberg}, {Hughes}, {De la Luz}, \&
  {Boley}}]{white20}
{White}, J.~A., {Tapia-V{\'a}zquez}, F., {Hughes}, A.~G., {et~al.} 2020, \apj,
  894, 76

\bibitem[{{Wright} {et~al.}(2004){Wright}, {Marcy}, {Butler}, \&
  {Vogt}}]{wright04}
{Wright}, J.~T., {Marcy}, G.~W., {Butler}, R.~P., \& {Vogt}, S.~S. 2004, \apjs,
  152, 261

\end{thebibliography}

\begin{appendix}

\section{Input table}

Table~\ref{tab_targets} lists the star in our sample as well as the input parameters used in our computations. It includes  all F, G, and K stars from \cite{theia17}. 

\onecolumn

\begin{landscape}
\begin{longtable}{llllllllllllll}
\caption{\label{tab_targets} Input parameters}\\
\hline
\# & Name 1 & Name 2   & B-V & TS & THEIA   & Dist. & Mass & Luminosity & T$_{\rm eff}$ & $\log R'_{HK\rm min}$  & $\log R'_{HK\rm max}$ & PHZ$_{\rm mid}$ \\
 & & & & & M$_{\rm lim}$ & & & & & &  \\
 & &   &   &  & (M$_{\rm Earth}$)  & (pc) & (solar) & (solar) & (K) & & &  (days) \\
\hline
      1 & $\alpha$ Cen. A & Gl559A &  0.71 & G2V &  0.23 &  1.35(***) &  1.12 &  1.52 & 5585(22) & -5.059(8) & -4.970(4) &   671.1\\
      2 & $\alpha$ Cen. B & Gl559B &  0.90 & K1V &  0.30 &  1.35(***) &  0.84 &  0.51 & 5248(18) & -4.920(4) & -4.940(4) &   355.2\\
      3 & $\epsilon$ Eridani & Gl144 &  0.88 & K2V &  1.08 &  3.22(*) &  0.76 &  0.35 & 4999(8) & -4.598(8) & -4.331(2) &   290.2\\
      4 & 61 Cygni A & Gl820A &  1.18 & K5V &  1.11 &  3.50(**) &  0.59 &  0.13 & 4379(24) & -4.910(2) & -4.668(2) &   172.4\\
      5 & 61 Cygni B & Gl820B &  1.37 & K7V &  1.25 &  3.49(**) &  0.52 &  0.08 & 4040(24) & -4.999(12) & -4.900(5) &   130.6\\
      6 & Procyon A & Gl280A &  0.42 & F5IV-V &  0.46 &  3.51(*) &  1.65 &  6.76 & 6592(22) & -4.820(2) & -4.614(2) &  1483.3\\
      7 & $\epsilon$ Indi & Gl845 &  1.06 & K5V &  1.32 &  3.64(**) &  0.65 &  0.19 & 4571(22) & -4.851(6) & -4.809(13) &   210.4\\
      8 & $\tau$ Ceti & Gl71 &  0.72 & G8V &  1.09 &  3.65(*) &  0.81 &  0.44 & 5338(8) & -5.065(8) & -4.895(2) &   322.2\\
      9 & Groombridge 1618 & Gl380 &  1.33 & K6V &  2.19 &  4.87(**) &  0.53 &  0.09 & 4002(24) & -4.682(12) & -4.615(2) &   137.1\\
     10 & 70 Ophiuchi A & Gl702A &  0.86 & K0V &  1.16 &  5.08(*) &  0.84 &  0.52 & 5023(22) & -4.663(2) & -4.367(2) &   370.3\\
     11 & 70 Ophiuchi B & Gl702B &  1.19 & K4V &  1.60 &  5.12(**) &  0.59 &  0.13 & 4475(24) & -4.759(3) & -4.557(1) &   169.1\\
     12 & $\sigma$ Draconis & Gl764 &  0.78 & K0V &  1.80 &  5.75(*) &  0.78 &  0.40 & 5272(22) & -4.921(12) & -4.666(2) &   304.7\\
     13 & 33G Librae A & Gl570A &  1.11 & K4V &  1.85 &  5.88(**) &  0.66 &  0.21 & 4519(22) & -4.875(16) & -4.480(4) &   221.7\\
     14 & $\eta$ Cassio. A & Gl34A &  0.58 & F9V+M0V &  1.19 &  5.95(*) &  1.05 &  1.22 & 5902(22) & -4.985(2) & -4.930(7) &   563.7\\
     15 & $\eta$ Cassio. B & Gl34B &  1.39 & K7V &  2.39 &  5.93(**) &  0.48 &  0.06 & 4011(23) & -4.936(2) & -4.806(12) &   112.2\\
     16 & 36 Ophiuchi A & Gl663A &  0.85 & K2V &  1.43 &  5.96(**) &  0.70 &  0.26 & 5370(18) & -4.684(2) & -4.406(2) &   233.0\\
     17 & 36 Ophiuchi B & Gl663B &  0.85 & K1V &  1.44 &  5.96(**) &  0.70 &  0.26 & 5370(18) & -4.664(2) & -4.415(2) &   228.9\\
     18 & 279G Sagit. A & Gl783 &  0.87 & K2.5V &  2.08 &  6.02(**) &  0.69 &  0.24 & 4857(8) & -5.079(8) & -4.940(9) &   232.9\\
     19 & 82G Eridani & Gl139 &  0.71 & G6V &  1.76 &  6.04(*) &  0.91 &  0.69 & 5478(8) & -5.025(13) & -4.949(15) &   418.7\\
     20 & $\delta$ Pavonis & Gl780 &  0.76 & G8IV &  1.43 &  6.11(*) &  1.03 &  1.13 & 5512(8) & -5.030(4) & -4.980(4) &   564.5\\
     21 & $\xi$ Bootis A & Gl566A &  0.73 & G7Ve &  1.59 &  6.73(**) &  0.85 &  0.54 & 5570(21) & -4.442(2) & -4.146(2) &   353.2\\
     22 & $\xi$ Bootis B & Gl566B &  1.17 & K5Ve &  2.37 &  6.75(**) &  0.54 &  0.10 & 4418(24) & -4.418(1) & -4.356(12) &   140.8\\
     23 & $\beta$ Hydri & Gl19 &  0.62 & G0V &  1.31 &  7.46(*) &  1.30 &  2.68 & 5784(8) & -5.050(4) & -4.960(4) &   928.7\\
     24 & $\mu$ Cassio. A & Gl53A &  0.69 & G5V &  2.16 &  7.55(*) &  0.79 &  0.42 & 5395(22) & -4.992(2) & -4.933(2) &   309.1\\
     25 & $\pi$ 3O Orionis & Gl178 &  0.44 & F6V &  1.58 &  8.07(*) &  1.25 &  2.37 & 6412(22) & -4.650(7) & -4.626(2) &   793.7\\
     26 & pEridani A & Gl66A &  0.86 & K2V &  1.93 &  8.19(**) &  0.67 &  0.22 & 5023(22) & -4.899(8) & -4.740(4) &   214.3\\
     27 & pEridani B & Gl66B &  0.90 & K2V &  2.00 &  8.19(**) &  0.64 &  0.19 & 5093(22) & -4.830(6) & -4.830(6) &   194.9\\
     28 & $\mu$ Herculis A & Gl695A &  0.75 & G5IV &  1.30 &  8.31(*) &  1.33 &  2.96 & 5562(25) & -5.193(2) & -5.043(12) &  1017.4\\
     29 & $\gamma$ Pavonis & Gl827 &  0.48 & F9V &  2.12 &  9.26(*) &  1.06 &  1.24 & 6205(8) & -4.811(15) & -4.491(6) &   545.6\\
     30 & $\zeta$ Tucanae & Gl17 &  0.57 & F9.5V &  1.91 &  8.59(*) &  0.97 &  0.89 & 5991(8) & -4.954(13) & -4.810(4) &   457.0\\
     31 & $\xi$ Ursae Major A & Gl423A &  0.55 & F8.5V &  1.79 &  8.73(***) &  1.15 &  1.69 & 5875(22) & -4.218(2) & -4.218(2) &   690.6\\
     32 & $\xi$ Ursae Major B & Gl423B &  0.65 & G2V &  1.95 &  8.73(**) &  1.05 &  1.19 & 5650(19) & -4.289(2) & -4.289(2) &   571.1\\
     33 & $\gamma$ Leporis A & Gl216A &  0.47 & F6V &  1.36 &  8.93(*) &  1.18 &  1.89 & 6372(8) & -4.966(12) & -4.770(4) &   694.4\\
     34 & $\gamma$ Leporis B & Gl216B &  0.94 & K3 &  2.21 &  8.90(**) &  0.69 &  0.24 & 4875(22) & -4.657(6) & -4.100(4) &   234.8\\
     35 & $\delta$ Eridani & Gl150 &  0.92 & K0+IV &  1.81 &  9.04(*) &  1.42 &  3.80 & 4866(8) & -5.232(2) & -5.039(2) &  1288.8\\
     36 & $\beta$ Com. Ber. & Gl502 &  0.59 & F9.5V &  2.08 &  9.13(*) &  1.04 &  1.15 & 5970(22) & -4.832(2) & -4.619(12) &   536.8\\
     37 & $\beta$ Canum Ven. & Gl475 &  0.61 & G0V &  2.13 &  8.44(*) &  1.06 &  1.25 & 5888(18) & -4.990(10) & -4.879(2) &   572.5\\
     38 & 66G Cen. A & Gl442A &  0.67 & G2V &  2.06 &  9.29(**) &  1.00 &  1.01 & 5688(8) & -5.031(11) & -4.874(14) &   514.9\\
     39 & $\zeta$ Herculis A & Gl635A &  0.64 & G0IV &  1.29 & 10.10(***) &  1.61 &  6.05 & 6026(18) & -5.193(2) & -5.090(14) &  1491.2\\
     40 & $\zeta$ Herculis B & Gl635B &  0.80 & K0V &  2.19 & 10.10(***) &  0.89 &  0.64 & 5300(20) & -4.830(17) & -4.830(17) &   407.8\\
     41 & $\beta$ Virginis & Gl449 &  0.55 & F9V &  2.00 & 10.93(*) &  1.34 &  3.04 & 6109(22) & -4.966(12) & -4.813(12) &   963.7\\
     42 & $\eta$ Bootis & Gl534 &  0.57 & G0IV &  1.58 & 11.40(*) &  1.63 &  6.46 & 6012(22) & -5.250(10) & -5.250(10) &  1555.8\\
     43 & $\gamma$ Virginis A & Gl482A &  0.36 & F1-F2V &  1.48 & 11.62(**) &  1.37 &  3.27 & 6808(22) & -5.039(14) & -4.479(16) &   920.2\\
     44 & $\beta$ Tri. Aus. A & Gl601A &  0.29 & F1V &  1.43 & 12.38(*) &  1.75 &  8.30 & 7109(8) & - & - &  1573.5\\
     45 & $\gamma$ Virginis B & Gl482B &  0.36 & F0mF2V &  1.44 & 12.74(**) &  1.41 &  3.65 & 6694(23) & -4.298(16) & -4.298(16) &   999.6\\
     46 & $\gamma$ Cephei & Gl903 &  1.04 & K1III-IV &  2.08 & 13.54(**) &  1.99 & 13.77 & 4786(18) & -5.315(2) & -5.256(12) &  2882.8\\
     47 & $\beta$ Aquilae A & Gl771A &  0.85 & G8IV &  1.73 & 13.70(*) &  1.59 &  5.87 & 5057(8) & -5.238(6) & -5.034(1) &  1649.0\\
     48 & $\alpha$ Fornacis A & Gl127A &  0.53 & F6V &  1.99 & 13.95(**) &  1.44 &  4.03 & 6258(8) & -4.990(10) & -4.901(6) &  1126.1\\
     49 & $\theta$ Bootis A & Gl549A &  0.51 & F7V &  1.92 & 14.53(*) &  1.33 &  2.92 & 6152(22) & -4.601(14) & -4.410(14) &   935.7\\
     50 & $\eta$ Cephei & Gl807 &  0.91 & K0IV &  2.08 & 14.27(*) &  1.71 &  7.71 & 4898(22) & - & - &  1988.6\\
     51 & $\tau$ Bootis A & Gl527A &  0.49 & F7IV-V &  2.17 & 15.62(*) &  1.25 &  2.32 & 6310(22) & -4.964(2) & -4.964(2) &   793.9\\
     52 & 10 Ursae Major A & Gl332 &  0.43 & F3V+G5V &  2.16 & 16.07(*) &  1.44 &  3.99 & 6516(22) & -4.639(14) & -4.547(14) &  1081.7\\
     53 & $\Psi$ Velorum A & Gl351A &  0.34 & F3V &  2.15 & 18.31(**) &  1.46 &  4.21 & 6837(8) & -4.669(14) & -4.362(11) &  1071.4\\
     54 & $\Psi$ Velorum B & Gl351B &  0.53 & F &  2.33 & 18.41(**) &  1.34 &  3.05 & 6148(27) & - & - &   960.2\\
     55 & $\delta$ Gemini A & Gl271A &  0.34 & F2V &  2.22 & 18.54(*) &  1.71 &  7.64 & 6902(22) & - & - &  1536.5\\
\hline
\end{longtable}
\tablefoot{B, V, and spectral types from CDS except the spectral type for $\zeta$ Herculis A \cite[][]{morel01} and $\Psi$ Velorum B (F star deduced from its B-V value). The THEIA detection limits are from \cite{theia17}, as are the masses and luminosities. The distances are derived from parallaxes from Hipparcos \cite[][]{vanleeuwen07}, indicated with a (*), from GAIA DR2 \cite[][]{gaia18}, indicated with (**), and from other references for a few stars indicated as  (***):  $\alpha$ Cen A and B \cite[][]{pourbaix16}; $\xi$ Ursae Major A (Gaia value for the B component); $\zeta$ Herculis A and B \cite[][]{theia17}. 
References for  $\log R'_{HK\rm min}$  and  $\log R'_{HK\rm max}$ are the following: (1) \cite{noyes84}; (2) \cite{duncan91}; (3) \cite{baliunas95}; (4) \cite{henry96};   (5) \cite{radick98};  (6) \cite{gray03};  (7) \cite{wright04};  (8) \cite{gray06};  (9) \cite{jenkins06} ; (10) \cite{hall07};  (11) \cite{schroder09};  (12) \cite{isaacson10};  (13) \cite{lovis11b};  (14) \cite{pace13};  (15) \cite{meunier17b};  (16) \cite{borosaikia18} ; (17) Deduced from the age \cite[][]{morel01} and relationship from \cite{mamajek08}. 
Reference for T$_{\rm eff}$ are the following: (18) \cite{allende99}; (19) \cite{cayrel01}; (20) \cite{morel01}; (21) \cite{valenti05}; (8) \cite{gray06}; (22) \cite{holmberg09}; (23) \cite{aleo17}; (24) \cite{houdebine19}; (25) \cite{marfil19}; (26) \cite{white20} ; (27) Derived from a law between B-V and T$_{\rm eff}$ established from data in \cite{gray03}.    }
\end{landscape}

\section{Stellar variability}

\subsection{Comparison with the solar case}


\begin{figure*}[h]
\includegraphics{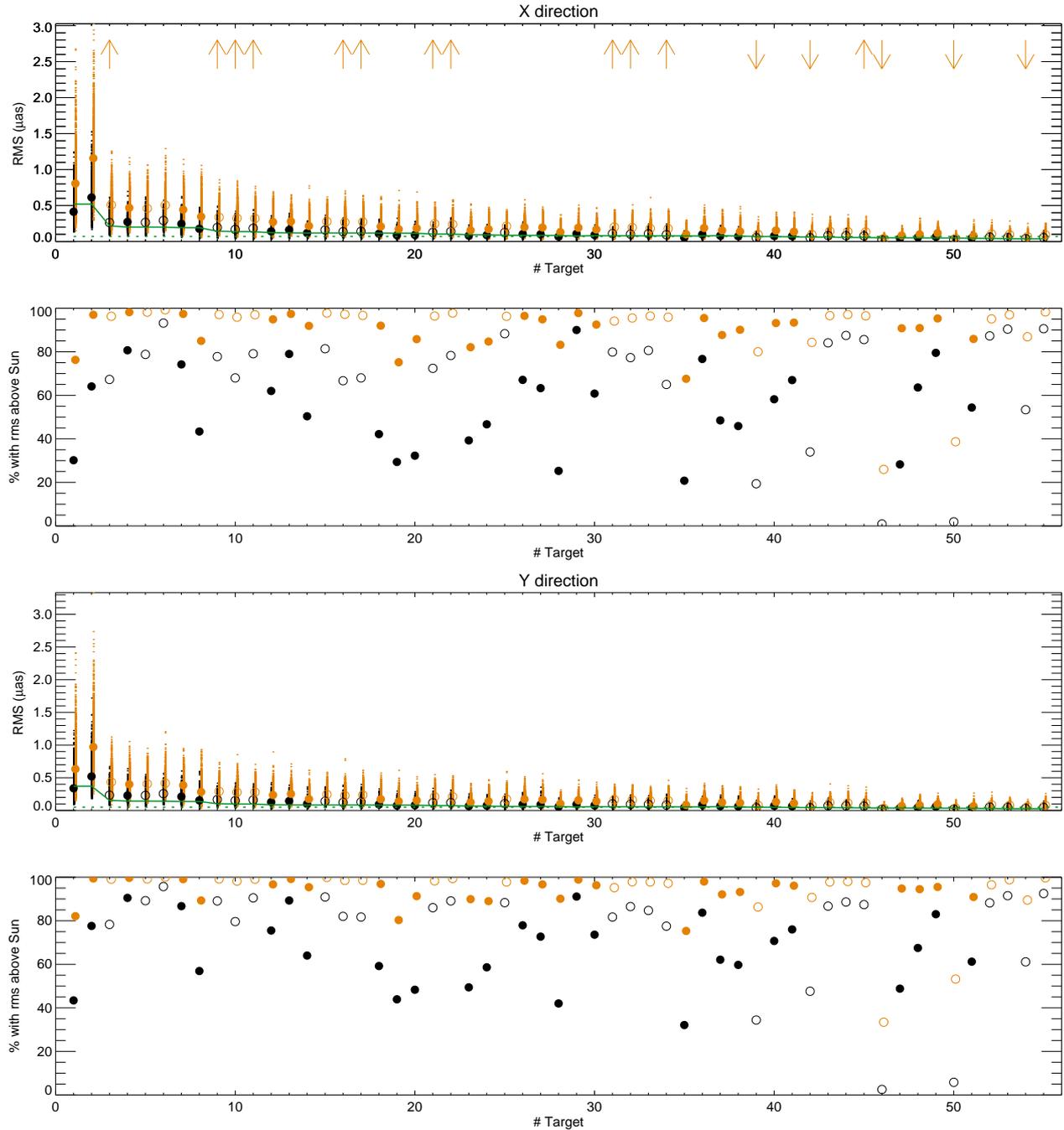}
\caption{Astrometric jitter and comparison to solar values. {\it First panel:} Astrometric jitter (rms) in the X  direction, for $\Delta T_{\rm spot1}$ (black dots) and $\Delta T_{\rm spot2}$ (orange dots). The median values are indicated by circles of the same colour (open circles correspond to stars outside the B-V range of our simulations). The green lines correspond to the solar value from \cite{lagrange11} at 10 pc (dotted line) and scaled to the distance of each star (solid line). Upward arrows mean that the rms are lower limits (active stars), while downward arrows correspond to upper limits (quiet stars, mostly subgiants). 
{\it Second panel:} Percentage of simulations with rms larger than the solar values of \cite{lagrange11} scaled to the distance of each star, for $\Delta T_{\rm spot1}$ (black) and $\Delta T_{\rm spot2}$ (orange). Open circles correspond to stars outside the B-V range of our simulations.
{\it Third panel:} Same as first panel for the Y direction.
{\it Fourth panel:} Same as second panel for the Y direction.
}
\label{rms}
\end{figure*}

\begin{table*}
\caption{Astrometric stellar variability}
\label{tab_var}
\begin{center}
\renewcommand{\footnoterule}{}  
\begin{tabular}{lllllllll}
\hline
\# & Name  & Number of  & Flag & Flag & rms X & rms Y & rms X & rms Y \\
    &               &        compatible    &  (Sp. Type) & (Activity)  &  ($\Delta$T$_{\rm spot1}$) &  ($\Delta$T$_{\rm spot1}$) &  ($\Delta$T$_{\rm spot2}$) &  ($\Delta$T$_{\rm spot2}$) \\
   &          &        simulations         & &   & ($\mu$as) &  ($\mu$as) &  ($\mu$as) &  ($\mu$as) \\
   \hline
      1 & $\alpha$ Cen. A &          243 & - & - &  0.41 &  0.81 &  0.34 &  0.63\\
      2 & $\alpha$ Cen. B &           81 & - & - &  0.61 &  1.16 &  0.52 &  0.97\\
      3 & $\epsilon$ Eridani &           81 & - & high &  0.26 &  0.51 &  0.23 &  0.43\\
      4 & 61 Cygni A &          243 & - & - &  0.27 &  0.47 &  0.23 &  0.40\\
      5 & 61 Cygni B &          162 & high B-V  & - &  0.27 &  0.46 &  0.23 &  0.41\\
      6 & Procyon A &          486 & low B-V  & - &  0.29 &  0.51 &  0.26 &  0.42\\
      7 & $\epsilon$ Indi &          162 & - & - &  0.25 &  0.44 &  0.21 &  0.39\\
      8 & $\tau$ Ceti &          405 & - & - &  0.18 &  0.35 &  0.15 &  0.28\\
      9 & Groombridge 1618 &           81 & high B-V  & high &  0.19 &  0.34 &  0.17 &  0.29\\
     10 & 70 Ophiuchi A &           81 & - & high &  0.17 &  0.32 &  0.15 &  0.28\\
     11 & 70 Ophiuchi B &           81 & - & high &  0.18 &  0.32 &  0.16 &  0.28\\
     12 & $\sigma$ Draconis &          486 & - & - &  0.14 &  0.27 &  0.13 &  0.24\\
     13 & 33G Librae A &          162 & - & - &  0.16 &  0.28 &  0.14 &  0.25\\
     14 & $\eta$ Cassio. A &          243 & - & - &  0.12 &  0.22 &  0.10 &  0.18\\
     15 & $\eta$ Cassio. B &          243 & high B-V  & - &  0.16 &  0.28 &  0.14 &  0.25\\
     16 & 36 Ophiuchi A &           81 & - & high &  0.14 &  0.27 &  0.12 &  0.23\\
     17 & 36 Ophiuchi B &           81 & - & high &  0.14 &  0.27 &  0.13 &  0.24\\
     18 & 279G Sagit. A &          324 & - & - &  0.11 &  0.21 &  0.09 &  0.17\\
     19 & 82G Eridani &          243 & - & - &  0.09 &  0.17 &  0.08 &  0.14\\
     20 & $\delta$ Pavonis &          243 & - & - &  0.09 &  0.18 &  0.08 &  0.15\\
     21 & $\xi$ Bootis A &           81 & - & high &  0.13 &  0.25 &  0.12 &  0.21\\
     22 & $\xi$ Bootis B &           81 & - & high &  0.14 &  0.24 &  0.12 &  0.21\\
     23 & $\beta$ Hydri &          243 & - & - &  0.08 &  0.16 &  0.07 &  0.12\\
     24 & $\mu$ Cassio. A &          243 & - & - &  0.09 &  0.17 &  0.08 &  0.14\\
     25 & $\pi$ 3O Orionis &          243 & low B-V  & - &  0.12 &  0.21 &  0.11 &  0.17\\
     26 & pEridani A &          324 & - & - &  0.10 &  0.20 &  0.09 &  0.17\\
     27 & pEridani B &          162 & - & - &  0.10 &  0.19 &  0.09 &  0.16\\
     28 & $\mu$ Herculis A &          162 & - & - &  0.07 &  0.13 &  0.06 &  0.10\\
     29 & $\gamma$ Pavonis &          486 & - & - &  0.11 &  0.19 &  0.09 &  0.16\\
     30 & $\zeta$ Tucanae &          405 & - & - &  0.09 &  0.17 &  0.08 &  0.14\\
     31 & $\xi$ Ursae Major A &           81 & - & high &  0.11 &  0.20 &  0.10 &  0.16\\
     32 & $\xi$ Ursae Major B &           81 & - & high &  0.10 &  0.19 &  0.09 &  0.16\\
     33 & $\gamma$ Leporis A &          405 & low B-V  & - &  0.11 &  0.18 &  0.09 &  0.15\\
     34 & $\gamma$ Leporis B &           81 & - & high &  0.09 &  0.18 &  0.08 &  0.15\\
     35 & $\delta$ Eridani &          162 & - & - &  0.05 &  0.10 &  0.04 &  0.08\\
     36 & $\beta$ Com. Ber. &          405 & - & - &  0.10 &  0.19 &  0.09 &  0.15\\
     37 & $\beta$ Canum Ven. &          324 & - & - &  0.08 &  0.15 &  0.07 &  0.12\\
     38 & 66G Cen. A &          405 & - & - &  0.07 &  0.14 &  0.06 &  0.11\\
     39 & $\zeta$ Herculis A &           81 & - & low  &  0.05 &  0.10 &  0.04 &  0.08\\
     40 & $\zeta$ Herculis B &          162 & - & - &  0.08 &  0.15 &  0.07 &  0.13\\
     41 & $\beta$ Virginis &          405 & - & - &  0.08 &  0.14 &  0.07 &  0.11\\
     42 & $\eta$ Bootis &           81 & - & low  &  0.05 &  0.10 &  0.04 &  0.08\\
     43 & $\gamma$ Virginis A &          648 & low B-V  & - &  0.08 &  0.14 &  0.07 &  0.12\\
     44 & $\beta$ Tri. Aus. A &          648 & low B-V  & all &  0.08 &  0.14 &  0.07 &  0.11\\
     45 & $\gamma$ Virginis B &           81 & low B-V  & high &  0.08 &  0.13 &  0.07 &  0.11\\
     46 & $\gamma$ Cephei &           81 & - & low  &  0.02 &  0.04 &  0.02 &  0.03\\
     47 & $\beta$ Aquilae A &          162 & - & - &  0.04 &  0.08 &  0.04 &  0.07\\
     48 & $\alpha$ Fornacis A &          243 & - & - &  0.06 &  0.10 &  0.05 &  0.08\\
     49 & $\theta$ Bootis A &           81 & - & - &  0.06 &  0.12 &  0.06 &  0.09\\
     50 & $\eta$ Cephei &           81 & - & low  &  0.02 &  0.04 &  0.02 &  0.04\\
     51 & $\tau$ Bootis A &           81 & - & - &  0.05 &  0.08 &  0.04 &  0.06\\
     52 & 10 Ursae Major A &          162 & low B-V  & - &  0.06 &  0.10 &  0.05 &  0.09\\
     53 & $\Psi$ Velorum A &          243 & low B-V  & - &  0.06 &  0.10 &  0.05 &  0.08\\
     54 & $\Psi$ Velorum B &           81 & - & low  &  0.04 &  0.07 &  0.03 &  0.06\\
     55 & $\delta$ Gemini A &          648 & low B-V  & all &  0.05 &  0.09 &  0.05 &  0.08\\
\hline
\end{tabular}
\end{center}
\tablefoot{The median of the rms are indicated in this table, and are based on the 1000 simulations for each target described in Sect.~\ref{sect233}, but no added noise (only the stellar contribution is considered here). 
}
\end{table*}

The astrometric jitter in each  direction is computed as the rms over each time series, with no additional noise. Figure~\ref{rms} shows the range of jitter over each of the 1000 time series for each target, in the X-direction (first panel) and Y-direction (third panel). The two colours correspond to the two possible choices for $\Delta T_{\rm spot}$. The solar value  from \cite{lagrange11} is indicated for comparison, both for the distance of 10 pc and adjusted to each target. The median of the rms for each target is given in Table~\ref{tab_var}. 
We have compared the obtained jitter with the solar value by computing the percentage of simulations above the solar value, which was for a Sun seen edge-on. These percentages are shown in the second and fourth panels of Fig.~\ref{rms}, and are usually above 50\% for $\Delta T_{\rm spot1}$, and above 80\% for $\Delta T_{\rm spot2}$. For example, in the X-direction, median percentage over the 55 stars of 78~\% for $\Delta T_{\rm spot1}$ and 97~\% for $\Delta T_{\rm spot2}$. This is in part due to the presence of more active simulations, to the distance, and to the larger spot contrast for $\Delta T_{\rm spot2}$ \cite[not considered in][which was for the solar case only]{lagrange11}. These large percentages therefore justify to reanalyse the effect of stellar activity on astrometric time series and planet detectability as performed in this paper.

\subsection{Stellar variability table}

Table~\ref{tab_var} provides the median of the stellar jitter in both directions and for both $\Delta T_{\rm spot}$ assumptions for all targets. It also indicates the correspondence with our large set of simulations: number of simulations which are compatible, and flags concerning the compatibility in terms of spectral type and average activity level (Sect.~\ref{sect233}). 


\

\section{Detection rates}
\label{appc}

\begin{table*}
\caption{Detection rates at THEIA detection limits}
\label{tab_rate}
\begin{center}
\renewcommand{\footnoterule}{}  
\begin{tabular}{llllllll}
\hline
\# & Name & SN$_{\rm peak}$ & SN$_{\rm peak}$ & 1\% Theo. & 1\% Theo. &  Blind test & Blind test  \\
 &      &                 rate        & rate &    rate        & rate  &   rate        & rate  \\
  &   &                  ( $\Delta$T$_{\rm spot1}$) &  ($\Delta$T$_{\rm spot2}$) & ($\Delta$T$_{\rm spot1}$) &  ($\Delta$T$_{\rm spot2}$) & ($\Delta$T$_{\rm spot1}$) &  ($\Delta$T$_{\rm spot2}$)  \\
  \hline
      1 & $\alpha$ Cen. A &  22.6 &   7.3 &  13.0 &   2.3 &  84.2 &  50.8\\
      2 & $\alpha$ Cen. B &   7.8 &   1.2 &  36.4 &   2.5 &  61.2 &  18.5\\
      3 & $\epsilon$ Eridani &  74.3 &  51.0 & 100.0 &  99.9 & 100.0 &  95.2\\
      4 & 61 Cygni A &  83.1 &  56.4 & 100.0 &  98.5 & 100.0 &  97.5\\
      5 & 61 Cygni B &  72.0 &  44.4 & 100.0 &  96.7 & 100.0 &  98.5\\
      6 & Procyon A &  24.6 &   7.0 &  99.6 &  93.8 & 100.0 &  94.2\\
      7 & $\epsilon$ Indi &  87.7 &  67.1 & 100.0 & 100.0 & 100.0 & 100.0\\
      8 & $\tau$ Ceti &  77.1 &  61.8 & 100.0 &  99.8 & 100.0 &  99.8\\
      9 & Groombridge 1618 &  91.4 &  79.3 & 100.0 & 100.0 & 100.0 & 100.0\\
     10 & 70 Ophiuchi A &  61.1 &  40.5 & 100.0 & 100.0 & 100.0 &  99.8\\
     11 & 70 Ophiuchi B &  61.3 &  47.4 & 100.0 &  98.2 &  96.2 &  87.8\\
     12 & $\sigma$ Draconis &  80.4 &  71.4 & 100.0 & 100.0 & 100.0 & 100.0\\
     13 & 33G Librae A &  87.0 &  73.2 & 100.0 & 100.0 & 100.0 & 100.0\\
     14 & $\eta$ Cassio. A &  65.8 &  55.0 & 100.0 & 100.0 & 100.0 & 100.0\\
     15 & $\eta$ Cassio. B &   5.1 &   4.8 &  32.5 &  34.2 &  33.0 &  33.5\\
     16 & 36 Ophiuchi A &  79.8 &  62.4 & 100.0 & 100.0 & 100.0 & 100.0\\
     17 & 36 Ophiuchi B &  77.8 &  63.4 & 100.0 & 100.0 & 100.0 &  99.8\\
     18 & 279G Sagit. A &  89.8 &  85.7 & 100.0 & 100.0 & 100.0 & 100.0\\
     19 & 82G Eridani &  79.6 &  76.4 & 100.0 & 100.0 & 100.0 & 100.0\\
     20 & $\delta$ Pavonis &  71.2 &  64.5 & 100.0 & 100.0 & 100.0 & 100.0\\
     21 & $\xi$ Bootis A &  68.4 &  57.0 & 100.0 & 100.0 & 100.0 & 100.0\\
     22 & $\xi$ Bootis B &  56.8 &  52.0 & 100.0 &  99.8 &  99.8 &  99.2\\
     23 & $\beta$ Hydri &  58.4 &  52.5 & 100.0 & 100.0 &  99.8 &  99.5\\
     24 & $\mu$ Cassio. A &  81.0 &  75.9 & 100.0 & 100.0 & 100.0 & 100.0\\
     25 & $\pi$ 3O Orionis &  67.2 &  55.3 & 100.0 & 100.0 & 100.0 & 100.0\\
     26 & pEridani A &  85.7 &  76.3 & 100.0 & 100.0 & 100.0 & 100.0\\
     27 & pEridani B &  88.6 &  80.7 & 100.0 & 100.0 & 100.0 &  99.8\\
     28 & $\mu$ Herculis A &  61.7 &  54.4 & 100.0 & 100.0 &  99.0 &  98.5\\
     29 & $\gamma$ Pavonis &  67.0 &  60.8 & 100.0 & 100.0 & 100.0 & 100.0\\
     30 & $\zeta$ Tucanae &  74.4 &  68.3 & 100.0 & 100.0 & 100.0 & 100.0\\
     31 & $\xi$ Ursae Major A &  73.1 &  62.7 & 100.0 & 100.0 & 100.0 & 100.0\\
     32 & $\xi$ Ursae Major B &  65.8 &  55.7 & 100.0 & 100.0 & 100.0 & 100.0\\
     33 & $\gamma$ Leporis A &  59.4 &  49.6 & 100.0 & 100.0 & 100.0 & 100.0\\
     34 & $\gamma$ Leporis B &  21.5 &  20.5 &  93.8 &  91.7 &  90.0 &  87.0\\
     35 & $\delta$ Eridani &  66.5 &  65.3 & 100.0 & 100.0 & 100.0 & 100.0\\
     36 & $\beta$ Com. Ber. &  71.0 &  62.0 & 100.0 & 100.0 & 100.0 & 100.0\\
     37 & $\beta$ Canum Ven. &  72.1 &  68.5 & 100.0 & 100.0 & 100.0 & 100.0\\
     38 & 66G Cen. A &  74.4 &  68.5 & 100.0 & 100.0 & 100.0 & 100.0\\
     39 & $\zeta$ Herculis A &  53.7 &  48.0 & 100.0 & 100.0 & 100.0 & 100.0\\
     40 & $\zeta$ Herculis B &  10.8 &   9.5 &  99.1 &  98.4 &  95.5 &  95.0\\
     41 & $\beta$ Virginis &  63.0 &  55.4 & 100.0 & 100.0 &  99.5 &  99.8\\
     42 & $\eta$ Bootis &  47.2 &  45.0 & 100.0 & 100.0 & 100.0 & 100.0\\
     43 & $\gamma$ Virginis A &  42.3 &  38.1 & 100.0 & 100.0 &  99.0 &  99.2\\
     44 & $\beta$ Tri. Aus. A &  40.5 &  37.5 & 100.0 & 100.0 & 100.0 &  99.8\\
     45 & $\gamma$ Virginis B &  20.8 &  20.7 & 100.0 & 100.0 &  99.0 &  99.0\\
     46 & $\gamma$ Cephei &  21.5 &  23.1 & 100.0 & 100.0 &  94.2 &  93.2\\
     47 & $\beta$ Aquilae A &  46.4 &  44.3 & 100.0 & 100.0 &  99.8 &  99.8\\
     48 & $\alpha$ Fornacis A &  60.4 &  52.3 & 100.0 & 100.0 &  99.2 &  99.8\\
     49 & $\theta$ Bootis A &  51.9 &  46.3 & 100.0 & 100.0 & 100.0 &  99.8\\
     50 & $\eta$ Cephei &  39.5 &  36.4 & 100.0 & 100.0 &  99.5 &  99.2\\
     51 & $\tau$ Bootis A &  58.3 &  53.1 & 100.0 & 100.0 & 100.0 & 100.0\\
     52 & 10 Ursae Major A &  55.8 &  47.7 & 100.0 & 100.0 &  99.0 &  98.5\\
     53 & $\Psi$ Velorum A &  46.8 &  44.4 & 100.0 & 100.0 &  98.0 &  97.8\\
     54 & $\Psi$ Velorum B &  17.5 &  15.4 & 100.0 & 100.0 &  99.0 &  98.0\\
     55 & $\delta$ Gemini A &  43.3 &  41.8 & 100.0 & 100.0 &  99.8 & 100.0\\
\hline
\end{tabular}
\end{center}
\tablefoot{Rates are in \% and computed for the detection limits from \cite{theia17}, corresponding to the middle of the habitable zone, for three methods: a threshold of 6 on SN$_{\rm peak}$ (Appendix~\ref{appc1}), theoretical false positive level (~1\%, Appendix~\ref{appc2}) and blind test with a fap of 1~\% (Sect.~\ref{sect3}). 
}
\end{table*}

\begin{figure*}[h]
\includegraphics{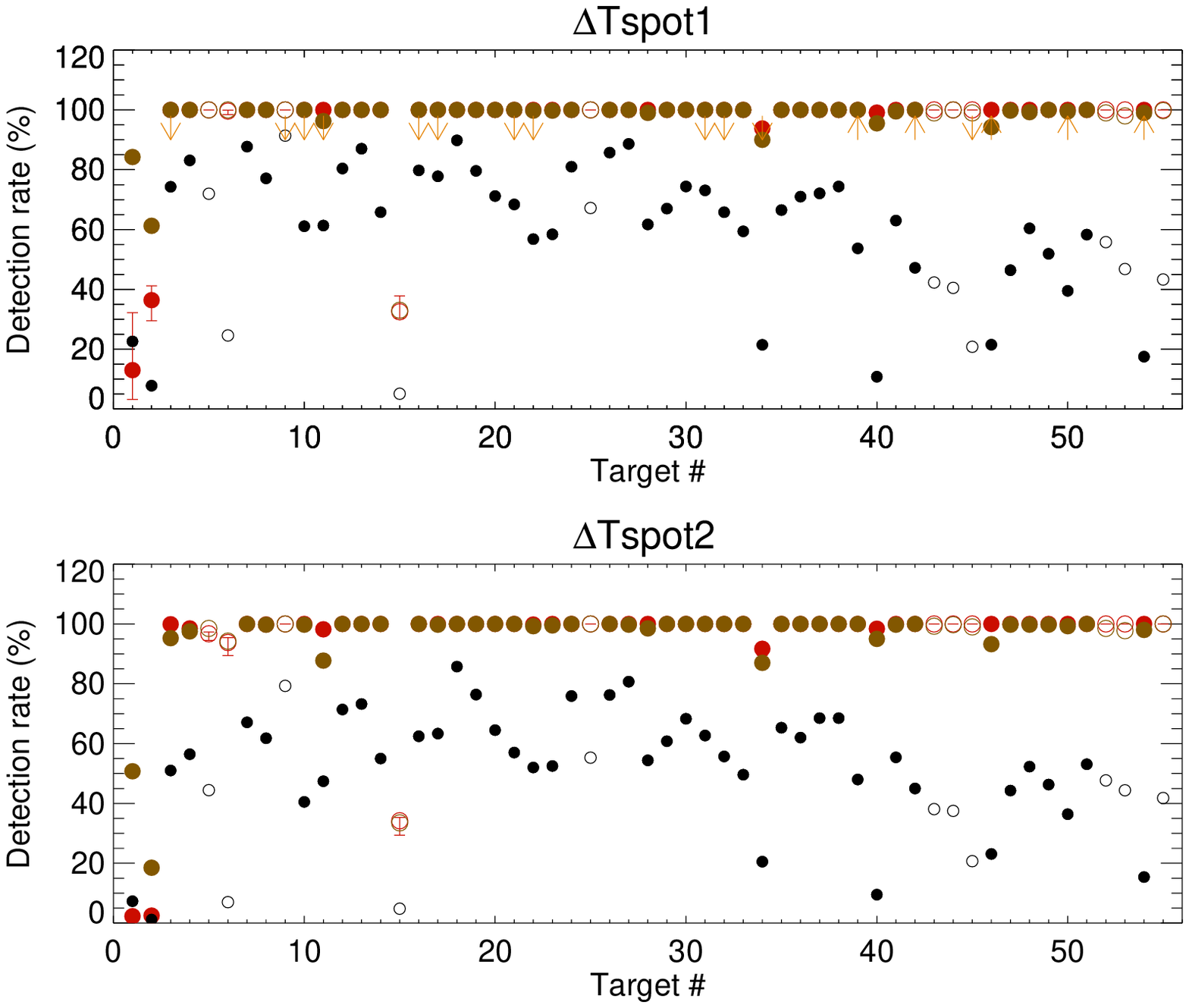}
\caption{Detection rates vs. target number for a planet mass at the detection limit of \cite{theia17}, for $\Delta T_{\rm spot1}$ (upper panel) and $\Delta T_{\rm spot2}$ (middle panel), and for different definitions: SN$_{\rm peak}$ $>$ 6 (black, Appendix~\ref{appc1}), false positive of 1~\% in the frequential analysis (red, Appendix~\ref{appc2}), and blind test with a fap at 1~\% (brown, Sect.~\ref{sect3}). Downward arrows mean that the detection rates are upper limits (active stars), while upward arrows correspond to lower limits (quiet stars, although most rates are already close to 100~\%). Open circles correspond to stars outside the B-V range of our simulations.
}
\label{tauxdet_all}
\end{figure*}

 In this section, we explore two complementary approaches to estimate the detection rates for the THEIA detection limit, first based on a definition of the S/N taking the frequency dependence into account, and then on a theoretical estimate of the false positive level. 
Table~\ref{tab_rate} lists the detection rates obtained with different methods for planets in the middle of the habitable zone and a mass corresponding to the detection limit from \cite{theia17}.

\subsection{Detection rates based on SN$_{\rm peak}$}
\label{appc1}

We have shown in Sect.~\ref{sect24} that the use of SN$_{\rm glob}$ for the detection limits in \cite{theia17} and our stellar estimation correspond to a relatively low level, significantly below 6. However, this definition does not take into account the fact that the planet and the noise (including the stellar noise) have a different frequential behaviour. Here, we therefore use another definition of the S/N, based on the peak in the  periodogram of the time series, hereafter SN$_{\rm peak}$, introduced in Paper II: the signal S is the amplitude of the planetary peak in the  periodogram, and the noise N is the maximum level of the  periodogram outside that peak and above 100 days.

Figure~\ref{sn} (lower panel) shows the typical values of SN$_{\rm peak}$ for a planet with the same mass, i.e. the detection limit from \cite{theia17} and in the middle of the habitable zone. It can be compared with SN$_{\rm glob}$ as is shown in the upper panel. We note that the distributions are asymmetrical, with an extension towards large S/N, as the 5th and 95th percentile levels are not symmetric with respect to the median. The median SN$_{\rm peak}$ over the 55 stars is 6.9 and 6.2, for $\Delta T_{\rm spot1}$ and $\Delta T_{\rm spot2}$ respectively, so quite close to the threshold of 6 targeted in \cite{theia17}. The median, however, varies from one star to the other, covering a range between 2.4 and 11.7 for $\Delta T_{\rm spot1}$ for example.

Using this definition and a threshold of 6, we can therefore compute the detection rate corresponding to the published detection limits by computing for each target the percentage of simulations (out of the 1000 synthetic time series) with SN$_{\rm peak}$ higher than 6. The results are shown in Figure~\ref{tauxdet_all} (black symbols). Most of them are in the 40--80~\% range. The median values over the 55 stars are given in  Table~\ref{tab_rate2}, and values for all stars are in Table~\ref{tab_rate}. Note that the detection rate is  small for $\alpha$ Cen A and B, which is explained in Sect.~\ref{sect4}.

\subsection{Detection rates based on theoretical false positives using a frequential approach}
\label{appc2}

In this second approach, we proceeded as in Paper II. We first computed the periodograms for the 1000 simulations of each target with no planet, and derived the maximum of the  periodogram around the period we are interested in (here the middle of habitable zone, considering the range 0.5P$_{\rm pla}$-2P$_{\rm pla}$). Finally, we established the false positive level, hereafter fp, such that 1~\% of the values are above, corresponding to a 1~\% false positive level. We then added on the same simulations a planet in the middle of the habitable zone and for a mass corresponding to the detection limit in \cite{theia17}. The maximum of the periodogram in the same frequency domain was then computed. The percentage of values above fp provides the detection rate.

The resulting detection rates are shown in Fig.~\ref{tauxdet_all} (red symbols). The median values over the 55 stars are given in  Table~\ref{tab_rate2}, and values for all stars are in Table~\ref{tab_rate}.  Most of them are excellent,  close to 100~\%. As before, the detection rates are low for $\alpha$ Cen A and B, this is explained in Sect.~\ref{sect4}. It is also low for target 15, which is a B component with a  noisy signal. Note that for stars with a period beyond the temporal span of 3.5 years, these results are to be used with caution: the expected planetary signal is strong, but not well characterised.

We recall that here, the false positive level was computed in the range around the planetary period because we wish to focus on the false positives in the habitable zone. We therefore estimated the level of false positives in this period domain. This therefore does not exclude the possible presence of false positives at  low periods. This will be taken into account in the blind test described in the following section: when considering the highest peak over the whole period range, it is indeed often at low period. This means that in practice most false positives come from low periods, and not from the habitable zone domain.

\section{New detection limits}

\begin{figure}[h]
\includegraphics{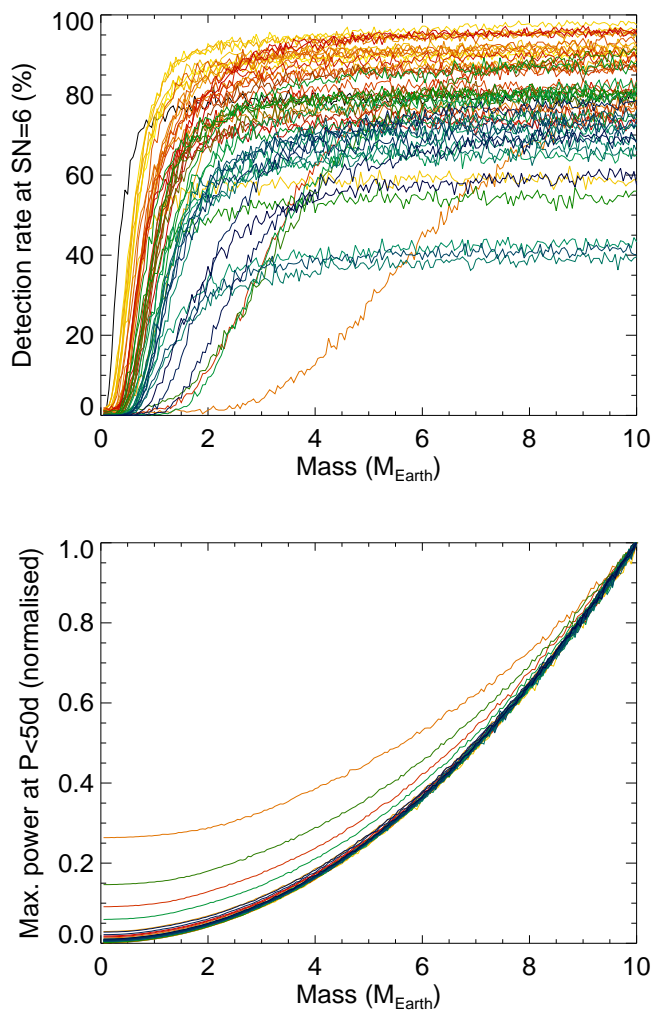}
\caption{Detection rates for SN$_{\rm peak}$ $>$ 6 versus planet mass (upper panel)  for all stars (one curve per star, arbitrary colours). The lower panel shows the maximum power for periods below 50 days vs. planet mass for all stars (arbitrary colours), after normalisation to a maximum of 1. 
}
\label{extaux}
\end{figure}

\begin{figure*}[h]
\includegraphics{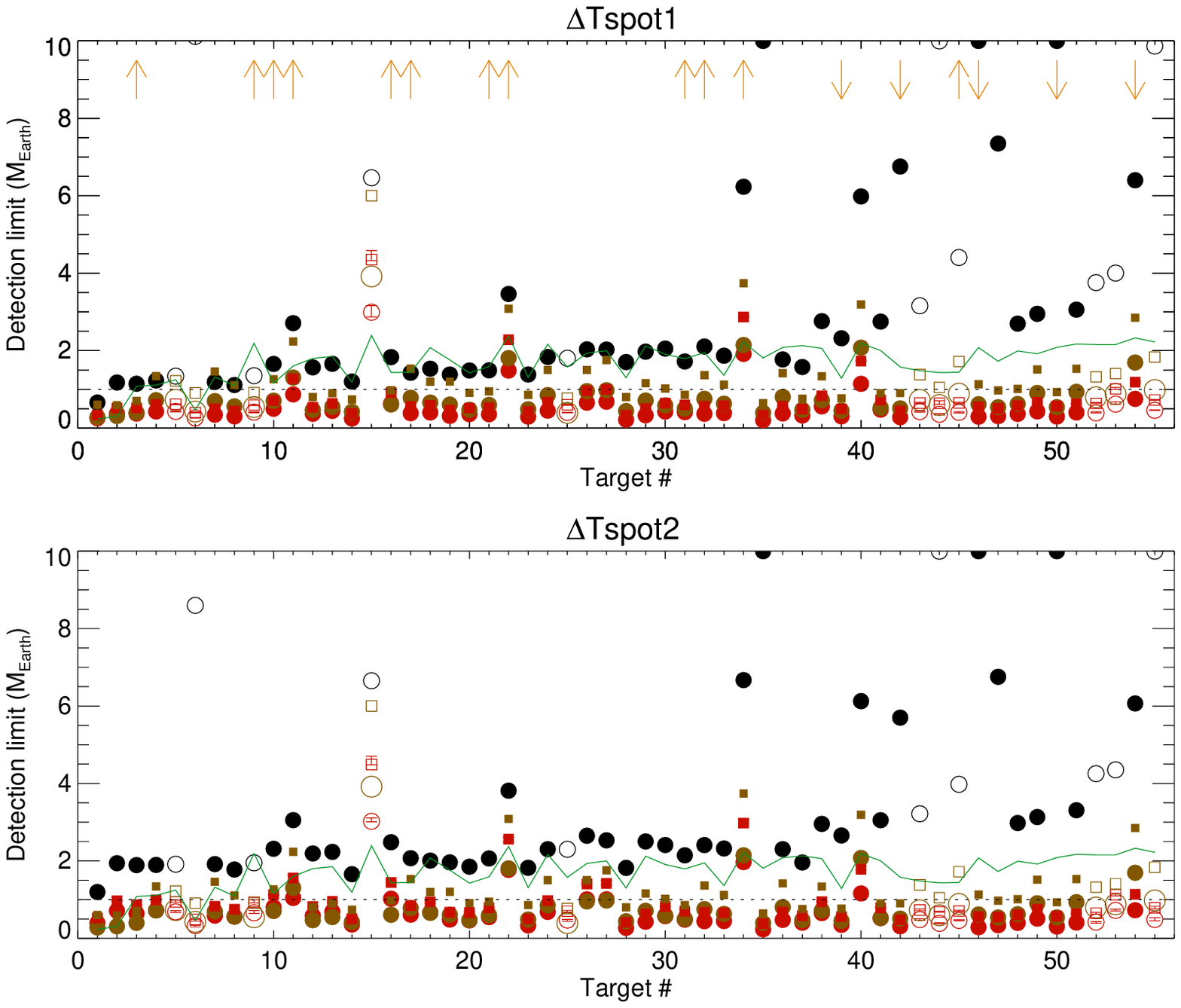}
\caption{
Detection limits vs. target number for $\Delta T_{\rm spot1}$ (upper panel) and $\Delta T_{\rm spot2}$ (lower panel), for different definitions: SN$_{\rm peak}$ $>$ 6 (black), theoretical fp of 1~\% (red, with uncertainty on the fp level, see text). The brown symbols are for blind test with a fap threshold of 1~\% made with a random selection between $\Delta T_{\rm spot1}$ and $\Delta T_{\rm spot2}$ on both plots. The  green line is the detection limit obtained in \cite{theia17}. Circles are for a detection rate threshold of 50~\%, and squares for a detection rate threshold of 95~\%. Open circles correspond to stars outside the B-V range of our simulations.
Downward arrows mean that the detection limits are upper limits (quiet stars), while upward arrows correspond to lower limits (quiet stars). Detection limits of 10 M$_{\rm Earth}$ are arbitrary (saturation of the detection rate curves, see Fig.~\ref{extaux}).
}
\label{limdet_all}
\end{figure*}

In this section, we explore two complementary approaches to estimate the new detection limits, first based on SN$_{\rm peak}$, and then on a theoretical estimate of the false positive level, as in Appendix~\ref{appc}. 
Table~\ref{tab_limdet} lists the detection limits obtained with different methods for planets in the habitable zone. Two different detection rates threshold (50~\% and 95~\%) are used. 

\subsection{Detection limits based on SN$_{\rm peak}$}
\label{appd1}

We considered planets with a random period within the habitable zone and performed a loop on the mass between 0.05 and 10 M$_{\rm Earth}$. The SN$_{\rm peak}$ was computed for all realisations of each target. The detection rate for each mass can then be computed by estimating the percentage of realisations with SN$_{\rm peak}$ higher than 6. We then estimated the detection limit for a given star as the mass for which the detection rate is 50~\%. 

Figure~\ref{extaux} (upper panel) shows all detection rates versus mass. 
Noticeably, the detection rate never reaches 100~\%, and it saturates at a different rate, depending on the star. Furthermore, the threshold of 50~\% cannot be applied for a few stars, as their saturation level being below 50~\%. The reason for this general behaviour is that when the planet mass is increased (and its contribution large compared to the noise and stellar contribution), the power of the planet peak is larger, but the same is true for the power in the periodogram at all periods, and in particular at low periods (far from the planetary regime considered here). Therefore, the S/N in the periodogram does not increase like the mass. An illustration is shown in the lower panel of Fig.~\ref{extaux}: The maximum power below 50 days is plotted versus the planet mass (after normalisation of all curves), showing a strong non-linear increase which is similar for most stars. 
This means that this S/N definition may not be adapted for large masses, since for those SN$_{\rm glob}$ would obviously  increase to larger amounts. SN$_{\rm peak}$ still leads to more optimistic detection limits however (see median in Table~\ref{tab_lim}): For comparison, the detection limit for a threshold of 6 applied to SN$_{\rm glob}$ has a higher median (4.1~M$_{\rm Earth}$ for $\Delta T_{\rm spot1}$ and 6.2~M$_{\rm Earth}$ for $\Delta T_{\rm spot2}$).  
It also means that there are planet masses corresponding to planet peaks far above the 1~\% false positive level with still a relatively low value of SN$_{\rm peak}$. 
Figure~\ref{limdet_all} shows the resulting detection limits (black circles). The detection limits for all stars are listed in Table~\ref{tab_limdet}.

\subsection{Detection limits based on theoretical false positives using a frequential approach}
\label{appd2}

\begin{figure}[h]
\includegraphics{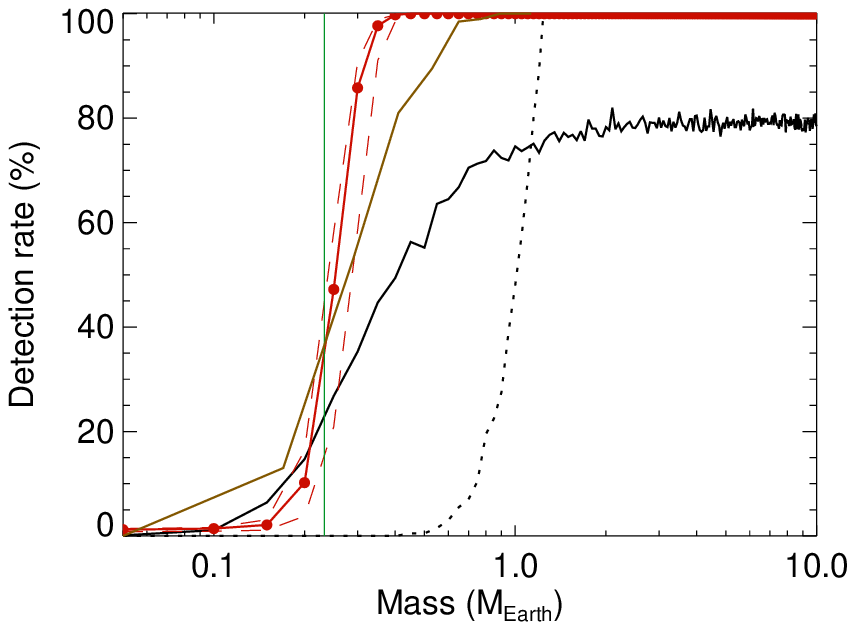}
\caption{Example of detection rates vs. mass for $\alpha$ Cen A for different definitions: blind test with a fap threshold of 1~\% (brown, all $\Delta T_{\rm spot}$, see Sect.~\ref{sect4}), SN$_{\rm peak}$ $>$ 6 (solid black line, $\Delta T_{\rm spot1}$, see Appendix~\ref{appd1}), SN$_{\rm glob}$ $>$ 6 (dahed black line, $\Delta T_{\rm spot1}$), and theoretical false positive of 1\% (in red, $\Delta T_{\rm spot1}$, see Appendix~\ref{appd2}). The dashed red line indicates the uncertainty on the red solid curve. The vertical green line is the detection limit obtained in \cite{theia17}.  
}
\label{extaux2}
\end{figure}

A similar loop on the mass is applied to the   periodogram computation and estimation of the planet power (period chosen randomly in the habitable zone), which is compared to the same false positive level fp as in Appendix~\ref{appc2} (i.e. computed with no planet). For a false positive level of 1~\%, the detection rates versus mass are used to compute the detection limits. An example is shown in Figure~\ref{extaux2} (red curve) for $\alpha$ Cen A. The curve is steeper than for SN$_{\rm peak}$ (black curve), increases for a lower mass, and reaches 100~\%. 
The range in mass between a 5~\% and a 95~\% detection rate is small, with a median value of 0.4~M$_{\rm Earth}$ for $\Delta T _{\rm spot1}$ (0.5~M$_{\rm Earth}$ for $\Delta T _{\rm spot2}$). This explains why in Appendix~\ref{appc2}, the detection rates for the detection limit of \cite{theia17} was  small for that star, although the mass is not far from this new detection limit.

Figure~\ref{limdet_all} shows the detection limits for this method, computed for two thresholds for the detection rate: 50~\% and 95~\%. The median of the ratio between the 95~\% detection limit and the 50~\% detection limit is 1.5. The median over all stars is listed in Table~\ref{tab_lim}, and Table~\ref{tab_limdet} provides  the values for the 55 stars. These detection limits correspond to a median SN$_{\rm peak}$ of 3.8 $\Delta T _{\rm spot1}$ (3.8 for $\Delta T _{\rm spot2}$).

\begin{table*}
\caption{New detection limits}
\label{tab_limdet}
\begin{center}
\renewcommand{\footnoterule}{}  
\begin{tabular}{lllllll}
\hline
\# & Name  & SN$_{\rm peak}$   & 1\% Theo.  & 1\% Theo.  &  Blind test  & Blind test  \\
 &   & (50~\%)  & (50~\%) &  (95~\%) &  (50~\%) &  (95~\%) \\
\hline
      1 & $\alpha$ Cen. A &  0.93 &  0.32 &  0.32 &  0.28 &  0.60\\
      2 & $\alpha$ Cen. B &  1.56 &  0.54 &  0.54 &  0.31 &  0.59\\
      3 & $\epsilon$ Eridani &  1.52 &  0.51 &  0.51 &  0.40 &  0.70\\
      4 & 61 Cygni A &  1.56 &  0.58 &  0.58 &  0.72 &  1.34\\
      5 & 61 Cygni B &  1.63 &  0.56 &  0.56 &  0.73 &  1.22\\
      6 & Procyon A &  9.36 &  0.29 &  0.29 &  0.43 &  0.91\\
      7 & $\epsilon$ Indi &  1.55 &  0.46 &  0.46 &  0.70 &  1.46\\
      8 & $\tau$ Ceti &  1.44 &  0.43 &  0.43 &  0.56 &  1.11\\
      9 & Groombridge 1618 &  1.65 &  0.54 &  0.54 &  0.54 &  0.91\\
     10 & 70 Ophiuchi A &  1.98 &  0.65 &  0.65 &  0.71 &  1.26\\
     11 & 70 Ophiuchi B &  2.88 &  0.96 &  0.96 &  1.30 &  2.24\\
     12 & $\sigma$ Draconis &  1.88 &  0.47 &  0.47 &  0.47 &  0.80\\
     13 & 33G Librae A &  1.95 &  0.56 &  0.56 &  0.55 &  0.90\\
     14 & $\eta$ Cassio. A &  1.43 &  0.30 &  0.30 &  0.42 &  0.74\\
     15 & $\eta$ Cassio. B &  6.56 &  3.01 &  3.01 &  3.92 &  6.00\\
     16 & 36 Ophiuchi A &  2.16 &  0.83 &  0.83 &  0.61 &  0.98\\
     17 & 36 Ophiuchi B &  1.75 &  0.50 &  0.50 &  0.78 &  1.53\\
     18 & 279G Sagit. A &  1.77 &  0.53 &  0.53 &  0.66 &  1.20\\
     19 & 82G Eridani &  1.67 &  0.40 &  0.40 &  0.61 &  1.20\\
     20 & $\delta$ Pavonis &  1.67 &  0.41 &  0.41 &  0.47 &  0.91\\
     21 & $\xi$ Bootis A &  1.78 &  0.45 &  0.45 &  0.60 &  0.95\\
     22 & $\xi$ Bootis B &  3.64 &  1.63 &  1.63 &  1.81 &  3.08\\
     23 & $\beta$ Hydri &  1.60 &  0.31 &  0.31 &  0.49 &  0.86\\
     24 & $\mu$ Cassio. A &  2.07 &  0.57 &  0.57 &  0.85 &  1.50\\
     25 & $\pi$ 3O Orionis &  2.05 &  0.44 &  0.44 &  0.39 &  0.77\\
     26 & pEridani A &  2.34 &  0.83 &  0.83 &  0.95 &  1.50\\
     27 & pEridani B &  2.28 &  0.84 &  0.84 &  0.98 &  1.76\\
     28 & $\mu$ Herculis A &  1.76 &  0.24 &  0.24 &  0.44 &  0.80\\
     29 & $\gamma$ Pavonis &  2.24 &  0.38 &  0.38 &  0.71 &  1.16\\
     30 & $\zeta$ Tucanae &  2.23 &  0.50 &  0.50 &  0.59 &  1.02\\
     31 & $\xi$ Ursae Major A &  1.93 &  0.45 &  0.45 &  0.50 &  0.86\\
     32 & $\xi$ Ursae Major B &  2.26 &  0.40 &  0.40 &  0.75 &  1.36\\
     33 & $\gamma$ Leporis A &  2.09 &  0.42 &  0.42 &  0.63 &  1.12\\
     34 & $\gamma$ Leporis B &  6.45 &  1.94 &  1.94 &  2.14 &  3.74\\
     35 & $\delta$ Eridani & 10.00 &  0.22 &  0.22 &  0.39 &  0.64\\
     36 & $\beta$ Com. Ber. &  2.04 &  0.43 &  0.43 &  0.80 &  1.42\\
     37 & $\beta$ Canum Ven. &  1.77 &  0.36 &  0.36 &  0.47 &  0.76\\
     38 & 66G Cen. A &  2.86 &  0.61 &  0.61 &  0.72 &  1.34\\
     39 & $\zeta$ Herculis A &  2.49 &  0.32 &  0.32 &  0.45 &  0.77\\
     40 & $\zeta$ Herculis B &  6.05 &  1.15 &  1.15 &  2.08 &  3.19\\
     41 & $\beta$ Virginis &  2.90 &  0.50 &  0.50 &  0.52 &  0.90\\
     42 & $\eta$ Bootis &  6.23 &  0.29 &  0.29 &  0.51 &  0.90\\
     43 & $\gamma$ Virginis A &  3.19 &  0.45 &  0.45 &  0.72 &  1.38\\
     44 & $\beta$ Tri. Aus. A & 10.00 &  0.37 &  0.37 &  0.59 &  1.05\\
     45 & $\gamma$ Virginis B &  4.19 &  0.43 &  0.43 &  0.89 &  1.72\\
     46 & $\gamma$ Cephei & 10.00 &  0.29 &  0.29 &  0.61 &  1.13\\
     47 & $\beta$ Aquilae A &  7.05 &  0.32 &  0.32 &  0.54 &  0.97\\
     48 & $\alpha$ Fornacis A &  2.84 &  0.38 &  0.38 &  0.62 &  1.01\\
     49 & $\theta$ Bootis A &  3.04 &  0.47 &  0.47 &  0.91 &  1.51\\
     50 & $\eta$ Cephei & 10.00 &  0.30 &  0.30 &  0.54 &  0.93\\
     51 & $\tau$ Bootis A &  3.18 &  0.40 &  0.40 &  0.93 &  1.53\\
     52 & 10 Ursae Major A &  4.00 &  0.40 &  0.40 &  0.80 &  1.32\\
     53 & $\Psi$ Velorum A &  4.18 &  0.67 &  0.67 &  0.88 &  1.41\\
     54 & $\Psi$ Velorum B &  6.23 &  0.74 &  0.74 &  1.69 &  2.85\\
     55 & $\delta$ Gemini A &  9.93 &  0.47 &  0.47 &  0.98 &  1.83\\
\hline
\end{tabular}
\end{center}
\tablefoot{Detection limits are in M$_{\rm Earth}$ for the whole habitable zone. They have been obtained using three methods. For the detection obtained with SN$_{\rm peak}$ (Appendix~\ref{appd1}) and the theoretical false positive (Appendix~\ref{appd2}), we provide the average between the values obtained for $\Delta$T$_{\rm spot1}$ and $\Delta$T$_{\rm spot2}$.The blind tests (Sect.~\ref{sect3}) have been made with a combination of simulation using $\Delta$T$_{\rm spot1}$ and $\Delta$T$_{\rm spot2}$. Values of 10~M$_{\rm Earth}$ are arbitrary (see text). 
}
\end{table*}

\end{appendix}

\end{document}